\documentclass[superscriptaddress,amsmath,amssymb,aps,article,pre]{revtex4-2}
\pdfoutput=1

\usepackage{mathtools}
\usepackage{graphicx}
\usepackage{dcolumn}
\usepackage{cancel}
\usepackage{ulem}
\usepackage{bm}
\usepackage{physics}

\newcolumntype{P}[1]{>{\centering\arraybackslash}p{#1}}

\newcommand{\eq}{\text{eq}}

\newcommand{\Fbeta}{F}
\newcommand{\Fop}{\Fbeta[\omega](x,x')}
\newcommand{\Rej}{R[\omega](x)}
\newcommand{\Kbeta}{K}
\newcommand{\Kop}{\Kbeta[\omega](x,x')}
\newcommand{\lambdanu}{\lambda_{\nu}[\omega] }
\newcommand{\phinu}{\phi_{\nu}[\omega](x)}
\newcommand{\IPR}{\text{IPR}}
\newcommand{\CDW}{\text{CDW}}
\newcommand{\cutoff}{\text{cutoff}}
\newcommand{\calF}{\mathcal{F}}
\newcommand{\opt}{\text{opt}}
\newcommand{\seq}{\text{seq}}

\usepackage{hyperref}
\usepackage{footmisc}
\usepackage{color}

\begin{document}

\title{On the optimal convergence rate for the Metropolis algorithm in one dimension}

\author{A. Patrón}%
 \email{apatron@us.es}
 \affiliation{Física Teórica, Universidad de Sevilla, Apartado de
  Correos 1065, E-41080 Sevilla, Spain}
\author{A. D. Chepelianskii}%
 \email{alexei.chepelianskii@universite-paris-saclay.fr}
 \affiliation{Laboratoire de Physique des Solides, Universit\'e Paris-Saclay, CNRS, 91405, Orsay, France }
 \author{A. Prados}%
 \email{prados@us.es}
 \affiliation{Física Teórica, Universidad de Sevilla, Apartado de
  Correos 1065, E-41080 Sevilla, Spain}
\author{E. Trizac}%
\affiliation{LPTMS, Universit\'e Paris-Saclay, CNRS, 91405, Orsay, France }
\affiliation{Ecole Normale Sup\'erieure de Lyon, F-69364 Lyon, France}
 \begin{abstract}
    We study the relaxation of the Metropolis Monte Carlo algorithm corresponding to a single particle trapped in a one-dimensional confining potential, with even jump distributions that ensure that the dynamics verifies detailed balance: in particular, how can one minimise the characteristic time for reaching the target equilibrium probability distribution function? To work out the corresponding optimal sampling method, we study the physical mechanisms that affect the dynamics. Previous work suggested that, for smooth jump distributions, the fastest convergence rate is obtained as a result of the competition between diffusive and rejection-dominated dynamics. In this work, we show that a new relevant physical regime comes into play for two-peaked jump distributions, where the relaxation dynamics is dominated neither by diffusion nor by rejection: the eigenmodes adopt an oscillatory form, reminiscent of charge density waves (CDW)---thus we term this new regime the CDW regime. Using a combination of numerical and analytical techniques, the parameter regions corresponding to diffusion, rejection, and CDW are characterised, as well as the transition lines between them. Therefore, a phase diagram is built, in which there are two order parameters characterising each of the phase transitions: the Inverse Participation Ratio (IPR) and the Fidelity---reminiscent of the quantum fidelity. The optimal convergence rate is located at the triple point of phase coexistence, where the transition lines (diffusive-rejection, diffusive-CDW, and CDW-rejection) intersect. 
    We then address the fundamental problem of finding the optimal jump distribution, without any assumption on its functional form.
    Our theoretical framework is checked against the numerical diagonalisation of the master equation for the case of harmonic confinement, which is the paradigmatic example of a convex potential with only one minimum.  We also briefly discuss more sophisticated attempts at optimising the convergence rate to equilibrium.
 \end{abstract}

\maketitle


\section{Introduction}\label{sec:intro}

Computational algorithms based on Monte Carlo techniques are widely known for their versatility in solving many scientific problems. For instance, in molecular simulations, they are essential for determining the equilibrium properties of physical systems involving many degrees of freedom with high accuracy~\cite{frenkel_understanding_2023}. Other branches of knowledge that range from the natural sciences of physics, chemistry, and biology \cite{newman_monte_1999,mode_applications_2011} to the fields of economy, machine learning, and archaeology \cite{glasserman_monte_2004,bishop_pattern_2006,gilks_markov_1995} benefit from these techniques when approaching problems at the deterministic and stochastic levels of description, hence the interest in the study of the convergence of such algorithms, as it provides useful insights on the optimal performance they may reach.

The focus of our work concerns the Metropolis-Hastings algorithm, or simply Metropolis~\cite{metropolis_equation_1953}, which belongs to the class of Markov Chain Monte Carlo techniques. These are based on creating a sequence of steps from a random walk that {makes it possible to sample a desired target probability distribution, which usually corresponds to the equilibrium distribution, provided that the number of steps is large enough---i.e.~in the long-time limit, as discussed below.} In the Metropolis algorithm, the random displacements $\eta$ in parameter space giving the new steps of the sequence are drawn from a so-called jump distribution $\omega(\eta)$---{also known as proposal distribution---that} is an even function of the displacements. Once a random displacement is chosen, the associated energy change $\Delta U$ is calculated and the step is accepted or rejected according to the Metropolis rule: if $\Delta U<0$, it is accepted with probability $p=1$; if $\Delta U>0$, it is accepted with probability $p=\exp(-\beta \Delta U)$, where $\beta$ is the inverse temperature. This rule verifies detailed balance, and therefore it ensures that the system equilibrates in the long-time limit~\cite{gardiner_stochastic_2009,van_kampen_stochastic_1992}---provided that the jump distribution allows for the sampling of all configurations, i.e. ergodicity is not broken~\footnote{In the physical literature, ergodicity is often identified with irreducibility; the stochastic matrix of a Markov chain is said to be irreducible if any final state can be reached from an arbitrary initial state by means of a chain of transitions with non-zero probability.}.

A key issue concerns the overall convergence time of the algorithm; it quantifies the time needed to attain the target distribution, which greatly impacts its performance{---or equivalently, the convergence rate, which measures how fast the algorithm attains such target distribution}. The longer the convergence time, the larger the error in the computed observables. On the one hand, if the typical size $a$ of the attempted displacements is too small, most of the jumps are accepted but the phase space is not explored sufficiently, thus involving a long convergence time. On the other hand, if the size $a$ of these displacements is too large, most jumps are rejected because they are likely to involve a high energy cost. This discussion suggests that there must exist an optimal amplitude $a_{\opt}$ that minimises the convergence time for the Metropolis algorithm. Although there have been many attempts to derive $a_{\opt}$ for specific models \cite{gelman_efficient_1996,gelman_weak_1997,schmon_optimal_2022}, the general rule of thumb corresponds to choosing $a$ so that the acceptance and rejection rates are comparable, i.e. the probability of accepting or rejecting the new step is close to $0.5$ \cite{frenkel_understanding_2023,krauth_statistical_2006,talbot_optimum_2003}. 

{In this work, we intend to go one step further. Rather than determining what is the optimal size $a_{\text{opt}}$ for a given jump distribution $\omega (\eta)$, our work is focused on studying what is the optimal jump distribution $\omega_{\text{opt}}(\eta)$ that minimises the convergence time. This problem can be viewed as engineering the jump distribution to accelerate the dynamics of the system, thus sharing the objectives of the general field of shortcuts, started in the quantum field of shortcuts to adiabaticity~\cite{guery-odelin_shortcuts_2019} and later transposed to classical and stochastic systems as swift state-to-state transformations~\cite{guery-odelin_driving_2023}.  Note that engineering the jump distribution could be viewed as analogous to engineering the potential, since changing the shape of the jump distribution varies the jump moments---which provide the coefficients in the Kramers-Moyal expansion of the master equation~\cite{gardiner_stochastic_2009,van_kampen_stochastic_1992}.}

{In order to tackle the aforementioned optimisation problem, it becomes necessary to study and understand the physical mechanisms affchangecting the relaxation dynamics of the system. }In previous work ~\cite{chepelianskii_metropolis_2023}, the relaxation dynamics of the Metropolis algorithm was studied for a particle trapped in a one-dimensional confining potential $U(x)$, and the jump distribution that optimises the convergence rate to equilibrium was investigated. For different potentials and choices of the jump distribution, it was shown that a critical jump length $a^*$ appears where a localisation transition occurs: going from $a < a^*$ to $a > a^*$ radically changes the relaxation behaviour. This localisation transition mainly affects the eigenmodes of the master equation characterising the algorithm: for $a < a^*$, the eigenmodes resemble the diffusive modes obtained at the limit $a \rightarrow 0^+$, while, for $a > a^*$, the eigenmodes become spatially localised at discrete points $x_0$, where the rejection probability is maximal. Interestingly, it was shown that $a_{\opt} = a^*$ for most of the cases considered, implying that the optimal convergence time is obtained as a balance between {diffusion and rejection}. 

Here, we show that the diffusion-rejection picture described above breaks down when two-peaked jump distributions are considered for one-dimensional potentials. In such a scenario, {although the localisation transition still occurs, the optimal convergence rate is attained instead at a point where a crossing between the leading eigenvalues of the dynamics takes place. Eigenvalue crossing constitutes a fundamental phenomenon that plays a critical role in explaining and predicting the behaviour of physical systems across various domains, both in quantum~\cite{boscain_motion_2010,wang_energy-level_2016} and statistical mechanics~\cite{teza_eigenvalue_2023}.} We focus our study on {such} phenomenon and {how it relates to the emergence of a} new regime, in which the spatial dependence of the eigenmodes of the Metropolis algorithm varies drastically to an oscillatory behaviour, reminiscent of the \textit{charge density wave} (CDW) transition in correlated electron systems~\cite{gruner_charge_1985,gruner_dynamics_1988,chen_charge_2016}. We will refer to this regime, which competes with both the diffusion and rejection ones, as the CDW regime. {For the sake of simplicity, we present the derivation of the novel CDW regime for the case of convex potentials with only one minimum, although the final result applies to a broad class of confining potentials in one dimension.}

{The first of} the main goals of this paper is to characterise the novel CDW regime and analyse its implications for the optimisation of the convergence rate of the Metropolis algorithm. To do so, we introduce two order parameters, the fidelity $\calF$ and the inverse participation ratio (IPR), which allow us to characterise the parameter regions dominated by diffusion, rejection, and CDW. A key finding of our work is that the optimal convergence rate is placed at the triple point of coexistence of the diffusive, rejection, and CDW phases. {As our second major goal, we employ a variational approach to tackle the functional problem of finding the optimal jump distribution $\omega_{\text{opt}}(\eta)$, and discuss additional strategies on how to further optimise the convergence rate.}

The structure of this paper is as follows. Section~\ref{sec: relaxation} presents the mathematical framework behind the Metropolis algorithm. It also provides a brief account of the diffusion-rejection picture from Ref.~\cite{chepelianskii_metropolis_2023} and discusses the {emergence of the eigenvalue crossing phenomenon}. Section~\ref{sec:CDW} is devoted to the derivation and characterisation of the CDW regime. In addition, we numerically check the emergence of {its corresponding} oscillatory modes. The phase diagram of the Metropolis algorithm is analysed in Sec.~\ref{sec:phase-diag}, {for a specific} bi-parametric family of jump distributions. Then, in Sec.~\ref{sec: optimality}, {we introduce the variational method employed for estimating the optimal jump distribution. Also, we employ this variational method to further corroborate numerically} that optimality is attained as the competition between the CDW, diffusion, and rejection regimes. Section~\ref{sec:further-opt} presents some preliminary ideas for {estimating the shape of the optimal jump distribution} of the Metropolis algorithm. Finally, we provide some concluding remarks in Sec.~\ref{sec: conclusions}. The appendix discusses some technicalities that are omitted in the main text.

\section{Relaxation dynamics for the Metropolis algorithm}\label{sec: relaxation}

In this work, we study the relaxation dynamics of a particle confined in a one-dimensional confining potential $U(x)$, with $x$ accounting for its position on the real line{, and additionally immersed in a thermal bath at temperature $T$}. At each discrete time step $n$, the particle's position evolves according to the so-called Metropolis rule
\begin{equation}\label{eq:metropolis-rule}
    x_n = 
    \begin{dcases}
         x_{n-1} + \eta_n, &   \text{with prob.} \ {p(x_n,x_{n-1}) \equiv \text{min}\left(1,e^{-\beta \Delta U(x_n,x_{n-1})}\right)}, \\
         x_{n-1}, &   \text{with prob.} \ {1-p(x_n,x_{n-1})},
    \end{dcases}
\end{equation}
where ${\Delta U(x_n,x_{n-1})} \equiv U(x_n) - U(x_{n-1})$ and $\beta \equiv (k_BT)^{-1}$ is the inverse temperature. Independent random variables $\eta_n$ are drawn from a normalised distribution $\omega(\eta)$, which is usually termed the jump distribution. {Note} that $\Delta x=x_n-x_{n-1}=\eta_n$ {for accepted jumps, while $\Delta x = 0$ otherwise}. We further assume the jump distribution to be symmetric, i.e. $\omega(-\eta)=\omega(\eta)$, which ensures that detailed balance holds.

At the ensemble level, the system is described via the position probability distribution {$P_n[\omega](x)$, which depends on the functional form of the jump distribution $\omega(\eta)$}. Its dynamical evolution is governed by the master equation
\begin{equation}\label{eq:master-equation}
    P_n[\omega](x) = \int_{-\infty}^{+\infty} dx' \ \Fop P_{n-1}[\omega](x'),
\end{equation}
in which 
\begin{equation}
    \Fop \equiv \delta (x-x') \Rej + \omega(x-x'){p(x,x')},
\end{equation}
is the (temperature-dependent \footnote{To simplify our notation, we do not show the explicit dependence on the temperature. In simple cases like the harmonic potential, the temperature can be absorbed in the definition of the unit of length.}) integral kernel of the master equation{---i.e. the Markov kernel of the corresponding Markov chain}, and
\begin{equation}\label{eq:rejection-prob}
    \Rej\equiv \int_{-\infty}^{+\infty} dx' \ \omega(x'-x) {\left[1-p(x',x)\right]}. 
\end{equation}
We remark that $\Rej$ gives the rejection probability at position $x$, which is a central quantity in this work. Our notation emphasises the (functional) dependence of $P_n$, $F$, and $R$ on the jump distribution $\omega (\eta)$, in addition to their spatial dependence. 

The master equation~\eqref{eq:master-equation} has the canonical equilibrium distribution 
\begin{equation}\label{eq:P-eq}
    P_{\eq}(x) = \frac{1}{Z}e^{-\beta U(x)}, \quad Z\equiv \int_{-\infty}^{+\infty}dx \ e^{-\beta U(x)},
\end{equation}
as a stationary solution. The Metropolis rule~\eqref{eq:metropolis-rule} ensures that the equilibrium distribution is always attained in the long-time limit $n\to\infty$, regardless of the shape of the initial probability distribution $P_0(x)$, i.e.
\begin{equation}
    P_{\infty}(x)\equiv \lim_{n\to\infty}P_n[\omega](x)=P_{\eq}(x),
\end{equation}
provided that the jump distribution allows for {reaching any final position $x$ from an arbitrary initial position $x'$, i.e. the stochastic matrix $F[\omega](x-x')$ is irreducible---in the physics literature, this is usually identified with ergodicity. Note that, since the rejection probability $R[\omega](x)\ne 0$, the stochastic matrix is always aperiodic, so irreducibility---or ergodicity---ensures the convergence to the equilibrium distribution in the long-time limit. 
}

\subsection{Parametric families of jump distributions}\label{subsec:parametric-families}
We stress that our analysis on the convergence rate of the Metropolis algorithm is general, encompassing a wide class of jump distributions $\omega(\eta)$. Still, for illustration purposes, we consider some specific choices---the following bi-parametric families of jump distributions---throughout this work:
\begin{itemize}
    \item The $(a,\sigma)$-family of Gaussian jump distributions
    \begin{equation}\label{eq:parametric-gaussian}
    \omega^{(a,\sigma)} (\eta) = C^{(a,\sigma)} \text{exp}\left(-\frac{(|\eta|-a)^2}{2\sigma^2}\right),
    \end{equation}
    where $C^{(a,\sigma)}$ is a normalisation constant, and $a,\sigma >0$.
    \item The $(a,\alpha)$-family of algebraic jump distributions
    \begin{equation}\label{eq:alpha-a-distribution}
        \omega^{(a,\alpha)}(\eta) = \frac{\alpha + 1}{2a^{\alpha + 1}}|\eta|^{\alpha} \theta(a-|\eta|),
    \end{equation}
    with $a,\alpha >0$. In the above, $\alpha$ accounts for the width of the distribution, $a$ stands for the position of its maximum, {and $\theta (x)$ corresponds to the Heaviside step-function; $\theta(x)=1$ for $x\ge 0$ and $\theta(x)=0$ for $x< 0$.}
    \item The $(a,b)$-family of box-like jump distributions:
    \begin{equation}\label{eq:a-b-distribution}
        \omega^{(a,b)}(\eta) = \frac{1}{4b}\,\theta(|\eta|-a+b)\,\theta(a+b-|\eta|), \quad a>b>0.
    \end{equation}
    This distribution has a simple physical interpretation: only jumps with $a-b<|\eta|<a+b$ are allowed, with a flat distribution thereof. 
\end{itemize}
{The reasons behind these specific choices of families of jump-distributions are twofold: On the one hand, the parameter $a$---common for the three families---accounts for the typical size of attempted displacements, thus allowing to interpolate between the diffusion and rejection regimes previously uncovered in Ref.~\cite{chepelianskii_metropolis_2023}. On the other hand,} all these families tend to a double Dirac-delta peak
\begin{equation}\label{eq:omega0-delta-peak}
    \omega_0(\eta)=\frac{1}{2}\Big[\delta (\eta-a)+\delta (\eta+a) \Big],
\end{equation}
when their respective widths tend to zero, i.e. when (i) $\sigma\to 0^+$ in Eq.~\eqref{eq:parametric-gaussian}, (ii) $\alpha\to +\infty$ in Eq.~\eqref{eq:alpha-a-distribution}, and (iii) $b\to 0^+$ in Eq.~\eqref{eq:a-b-distribution}. The double delta peak does not allow for the sampling of all positions, since the length of the jumps is fixed to $\pm a$. This will be important in connection with the {eigenvalue crossing} discussed at the end of Sec.~\ref{sec:localisation-and-collapse}. {It is worth mentioning that the considered families of jump distributions are distinct enough such that they provide a sense of generality to the results that we present in the forthcoming sections. Specifically, while $\omega^{(a,\sigma)}(\eta)$ and $\omega^{(a,\alpha)}(\eta)$ have well-defined peaks at $\eta = \pm a$, $\omega^{(a,b)}(\eta)$ does not, as it presents a continuous of maxima in between the interval $a-b \leq \eta \leq a+b$. Moreover, while $\omega^{(a,\sigma)}(\eta)$ and $\omega^{(a,b)}(\eta)$ are symmetric with respect to $\eta = \pm a$, $\omega^{(a,\alpha)}(\eta)$ does not, as it presents an algebraic tail for $|\eta|\leq a$ and it vanishes for $|\eta|\geq a$.}

\subsection{Convergence rate}\label{sec:convergence-rate}

To study the convergence rate to equilibrium, we put our focus on the spectrum of eigenvalues of the integral kernel $\Fop$ of the master equation~\eqref{eq:master-equation}. For that purpose, it is useful to introduce a new integral kernel $\Kop$ that share the spectrum of eigenvalues with $\Fop$,
\begin{equation}\label{kernel-2}
    \Kop \equiv e^{-\beta {\Delta U(x,x')}/2}\Fop = \omega(x-x')e^{-\beta |{\Delta U(x,x')}|/2}+ \delta(x-x')\Rej.
\end{equation}
Moreover, $\Kop$ is symmetric under the exchange of $x\leftrightarrow x'$---note the absolute value in the exponential factor multiplying $\omega(x-x')$, which makes it self-adjoint and simplifies our mathematical description below.

Now we define the eigenvalues $\lambda_{\nu}$ and eigenfunctions $\phi_{\nu}(x)$ of the integral operator $\hat{K}$, 
\begin{equation}\label{eq:eigenproblem}
    \hat{K}[\omega] \, \phi_{\nu}(x)\equiv\int_{-\infty}^{+\infty}dx' \ \Kop\, \phi_{\nu}(x') = \lambdanu\, \phinu, \quad \nu=0,1,2,\ldots
\end{equation}
Since $\Kop$ is self-adjoint, all the eigenvalues of Eq.~\eqref{eq:eigenproblem} are real. Moreover, for jump distributions $\omega$ such that the Perron-Frobenius theorem \cite{bapat_nonnegative_1997} applies, this theorem ensures that (i) the maximum eigenvalue is $\omega$-independent, $\lambda_0=1$, which corresponds to the stationary solution being non-degenerate, and (ii) the remaining eigenvalues $\lambda_{\nu}[\omega]$, with $\nu>0$, verify the strict inequality $|\lambda_{\nu}|<1$. The eigenfunction $\phi_0(x)$ for the eigenvalue $\lambda_0=1$, corresponding to the equilibrium distribution, is also $\omega$-independent,
\begin{equation}
    \phi_0(x)\propto e^{-\beta U(x)/2}.
\end{equation}
Explicitly, we can write
\begin{subequations}\label{eq:eigenproblem-2}
\begin{equation}
    \int_{-\infty}^{+\infty}dx' \ \Kop\phi_0(x') = \phi_0(x), 
    \label{eq:eigenproblem-2-a}
    \end{equation}
    \begin{equation}
    \int_{-\infty}^{+\infty}dx' \ \Kop\phi_{\nu}[\omega](x') = \lambdanu\,\phinu, \quad \nu>0.
    \label{eq:eigenproblem-2-b}
\end{equation}
\end{subequations}
Our notation emphasises the fact that, under the conditions of the Perron-Frobenius theorem, both $\lambda_0$ and $\phi_0$ are independent of the jump distribution $\omega$, whereas the remainder of the eigenvalues and eigenfunctions are functionals thereof. Making use of the explicit form of the kernel $\Kop$ in Eq.~\eqref{kernel-2}, we have that
\begin{equation}
\int_{-\infty}^{+\infty}dx' \ \omega(x-x')e^{-\beta |{\Delta U(x,x')}|/2} \phi_{\nu}(x') +R[\omega](x)\, \phinu  = \lambdanu\,\phinu, \quad \nu>0.
    \label{eq:eigenproblem-2-b-explicit}
\end{equation}

The solution of the master equation can always be written as a linear combination of the eigenfunctions $\phi_{\nu}(x)$:
\begin{equation}
    P_n[\omega](x) = P_{\eq}(x)+e^{-\beta U(x)/2} \sum_{\nu>0}\mathcal{A}_{\nu}[\omega] \, \phinu \, (\lambda_{\nu}[\omega])^n,
\end{equation}
where $\mathcal{A}_{\nu}[\omega]$ are constants determined by the initial condition $P_0(x)$. For long times, the system always reaches equilibrium because $|\lambda_{\nu}[\omega]|<1$ for $\nu>0$. The leading convergence rate to equilibrium of the Metropolis algorithm {is thus dominated by the contribution} corresponding to the next-to-largest eigenvalue $\lambda_1[\omega]$, since
\begin{equation}\label{eq:Pn-long-times}
    P_n[\omega](x) -P_{\eq}(x) \sim e^{-\beta U(x)/2} \mathcal{A}_1[\omega] \, \phi_1[\omega](x) \, (\lambda_1 [\omega])^n, \quad n\gg 1.
\end{equation}

The next-to-largest eigenvalue can always be written as
\begin{equation}
\label{functional-true}
    \Lambda[\omega] \equiv \lambda_1[\omega] =  \underset{\phi \perp \phi_0}{\max}\ \int dx \int dx' \ \phi(x)\Kop\phi(x').
\end{equation}
{Equation~\eqref{eq:Pn-long-times} entails that $P_n[\omega](x) -P_{\eq}(x)\propto e^{-n|\ln\Lambda [\omega]|}$, so we can define the convergence rate as
\begin{equation}\label{eq:gamma-def}
    \gamma[\omega]\equiv |\ln\Lambda [\omega]|, \qquad P_n[\omega](x) -P_{\eq}(x)\propto e^{-n\,\gamma[\omega]}, \quad n\gg 1.
\end{equation}
The larger $\gamma[\omega]$, the faster the system equilibrates. Note that $\gamma[\omega]$ increases as $\Lambda[\omega]$ decreases, i.e. $\gamma[\omega]$ is an increasing function of the spectral gap $1-\Lambda[\omega]$ between the largest and next-to-largest eigenvalues.
} Our final goal is to select the optimal $\omega^*(\eta)$ that minimises $\Lambda[\omega]$, {i.e. maximises $\gamma[\omega]$ or the spectral gap $1-\Lambda[\omega]$,} in order to find the optimal convergence rate to equilibrium:
\begin{equation}\label{lambda*-def}
    \Lambda^*=\min_{\omega}\Lambda[\omega] {\implies \gamma^*=|\ln\Lambda^*|.}
\end{equation}
Therefore, we are actually dealing with a min-max problem. The minimisation in Eq.~\eqref{lambda*-def} should ideally be carried out over the whole space of possible jump distributions but, for practical purposes, it may also be performed within a given family of jump distributions, such as those in Eqs.~\eqref{eq:parametric-gaussian}-\eqref{eq:a-b-distribution}.

\subsection{Localisation transition and {eigenvalue crossing}}\label{sec:localisation-and-collapse}

In order to get more information about the physics of the problem, a sufficiently narrow jump distribution $\omega(\eta) = \delta_{\sigma}(\eta)$ around $\eta=0$ with standard deviation $\sigma$ was considered in Ref.~\cite{chepelianskii_metropolis_2023}. In this way, Eq.~\eqref{eq:eigenproblem} could be expanded around $x' = x$ due to the smallness of $\sigma$, giving as a result a Schrödinger-like equation with an effective potential,
\begin{equation}
    \sigma^2\left\{ - \frac{\phi_{n}''(x)}{2} +\left[ \frac{\beta^2}{8}(U'(x))^2- \frac{\beta}{4}U''(x)\right]\phi_{n}(x)\right\} = (1-\lambda_n)\phi_{n}(x).
\end{equation}
The eigenmodes $\phi_{n}(x)$ are the eigenfunctions of this Schrödinger-like equation, which have a well-defined parity for even potentials, such that $U(x)=U(-x)$. This makes it possible to split this Schrödinger basis into two parts that contain, respectively, the even and odd eigenfunctions. {This fact will become useful later on, specifically in Secs.~\ref{sec: optimality} and \ref{sec:further-opt}.}

In Ref.~\cite{chepelianskii_metropolis_2023}, it was also argued that the convergence rate is limited by the rejection probability $R[\omega](x)$. The main idea is the following: starting from an initial distribution $\delta(x-x_0)$ at step $n=0$, the probability distribution function at any later time is
\begin{equation}\label{eq:localised-modes}
    P_n(x) = R[\omega](x_0)^n\delta (x-x_0) + p_n(x|x_0),
\end{equation}
where $p_n(x|x_0)$ is a smooth function. The system cannot relax faster than this exponential decay, i.e., {the convergence rate should always be slower than the maximum of $|\ln R[\omega](x_0)|$ as a function of $x_0$. Equivalently, we can write
\begin{equation}\label{eq:lambda-ineq}
    \Lambda \geq \underset{x}{\max}\ R[\omega](x).
\end{equation}
}
  
The main result of Ref.~\cite{chepelianskii_metropolis_2023} is {that, for many different choices of jump distributions $\omega(\eta)$ and confining potentials $U(x)$---even at higher dimensions and including interactions, the optimal convergence rate is obtained} at the transition point between the diffusion and rejection regimes, which is where the \textit{localisation transition} takes place{---and where the inequality \eqref{eq:lambda-ineq} saturates}. The term localisation stems from the fact that, in addition to the discrete spectrum of well-defined eigenfunctions, a continuum of eigenvalues appears at the rejection regime, with singular localised eigenfunctions---as suggested by Eq.~\eqref{eq:localised-modes}, these singular eigenfunctions are peaked at any point $x_0$, with eigenvalue $R[\omega](x_0)$. Here we show that, although the above diffusion-rejection picture is quite robust for a wide class of jump distributions, this relatively simple scenario fails when considering two-peaked jump distributions---regardless of the chosen confining potential. 
\begin{figure}
  \centering
  \includegraphics[width=3.1in]{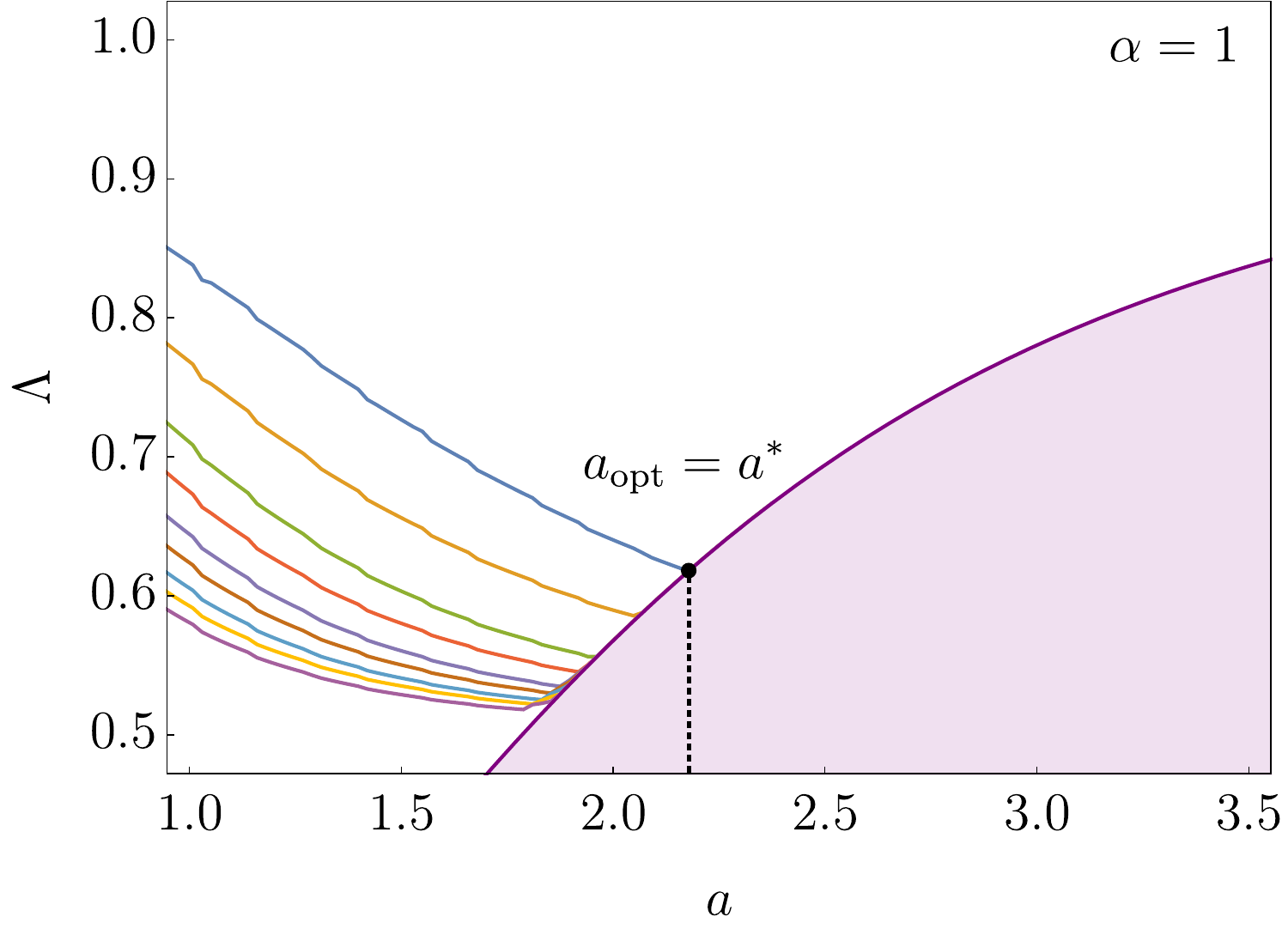}
  \includegraphics[width=3.1in]{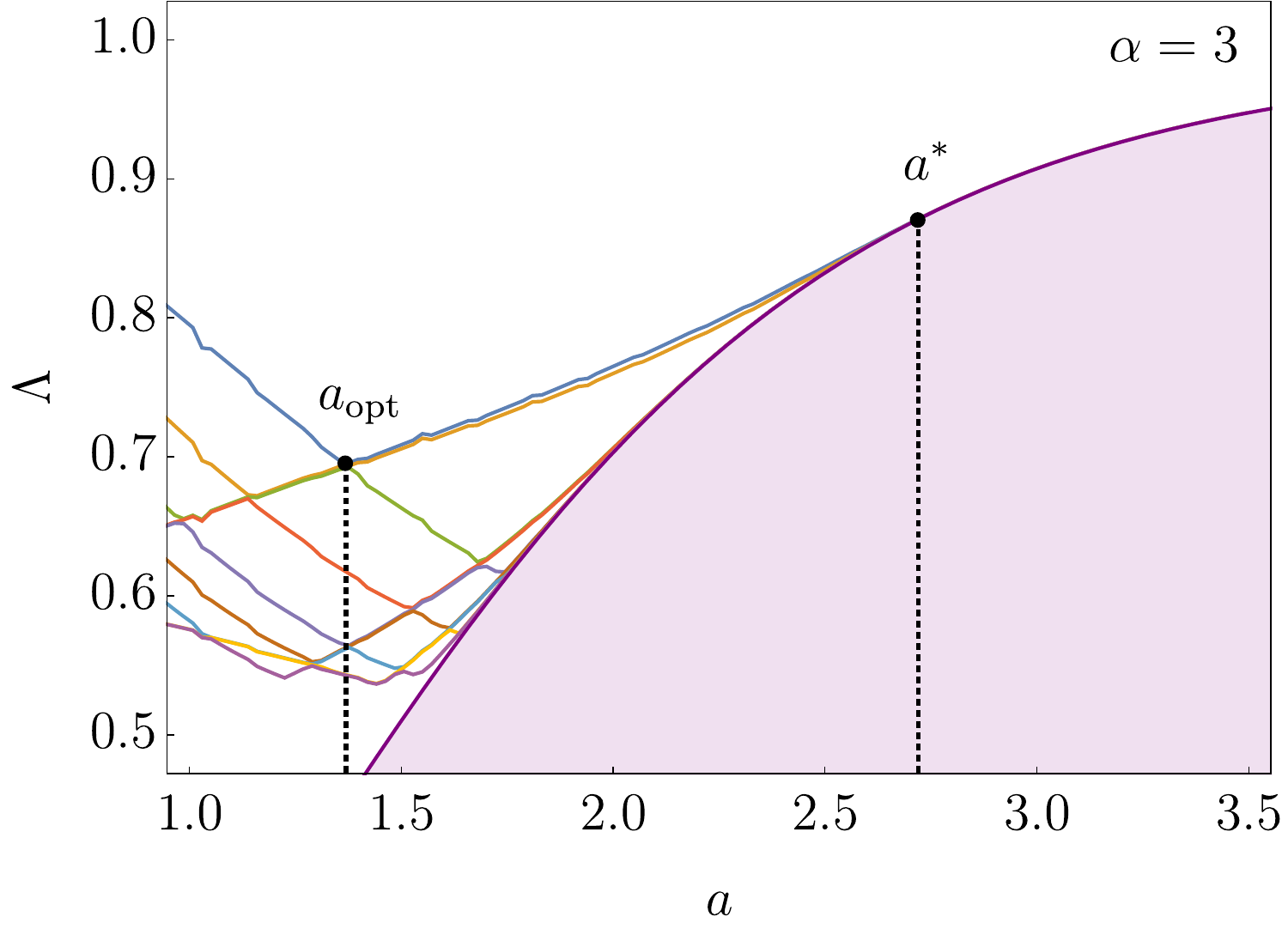}
  \caption{Leading eigenvalues $\lambda_{\nu}$, $\nu=1,\ldots,9$, of the master equation~\eqref{eq:master-equation} for the Metropolis algorithm, for a particle confined in a harmonic potential $U(x)=\chi x^2/2$. In both panels, we have employed the jump distribution $\omega^{(a,\alpha)}(\eta)$ from Eq.~\eqref{eq:alpha-a-distribution}, and we have plotted the eigenvalues against $a$ for two fixed values of $\alpha$: (left) $\alpha = 1$ and (right) $\alpha = 3$. Purple curves correspond to the maximal rejection probability, with the light purple regions accounting for the continuum of eigenvalues obtained upon the localisation transition. The dashed, vertical lines denote the position of either the optimal value $a_{\text{opt}}$ and/or the point $a^*$ at which the localisation transition takes place.
  }
  \label{fig_eigenvalue-crossing}
\end{figure}

Let us consider the above optimisation problem for a particle confined in a harmonic potential, {$U(x) = \chi x^2/2$---with $\chi$ being the stiffness, and the ($a,\alpha$)-family from Eq.~\eqref{eq:alpha-a-distribution}. As $\alpha$ increases, this family of jump distributions varies from a flat distribution in the interval $(-a,a)$, for $\alpha\ll 1$, to a two delta-peak distribution at $\eta=\pm a$, for $\alpha\gg 1$. In Fig.~\ref{fig_eigenvalue-crossing}, we plot the leading eigenvalues of the Metropolis algorithm $\lambda_{\nu}$, $1\le\nu\le 9$, as functions of $a$ for two specific values of the parameter $\alpha$, namely $\alpha=1$ and $\alpha=3$. For $\alpha = 1$, the left panel agrees with the diffusion-rejection picture developed in Ref.~\cite{chepelianskii_metropolis_2023}: the optimal value $a_{\text{opt}}$ is obtained at the point $a^*$ where the localisation transition takes place. However, this picture breaks as we increase the value of $\alpha$, and the jump distribution develops a two-peak structure. For $\alpha = 3$, we show in the right panel that $a_{\text{opt}}\neq a^*$: the optimal convergence rate is not attained at the localisation transition, but at a lower value where two eigenvalues cross and thus $\Lambda$ takes its minimum possible value. Eigenvalue crossing is related to abrupt changes in the dynamical behaviour of physical systems, thus pointing towards the emergence of new dominant physics---as shown, for instance, in Ref.~\cite{teza_eigenvalue_2023} in connection with dynamical phase transitions and the Mpemba effect~\cite{lu_nonequilibrium_2017,lasanta_when_2017,patron_non-equilibrium_2023}. In order to tackle the global problem of {maximising} the convergence rate of the Metropolis algorithm, it is important to understand all the physical mechanisms that come into play in its relaxation dynamics. The latter constitutes the first main goal of our work, which we next explore in Secs.~\ref{sec:CDW} and ~\ref{sec:phase-diag}.}

\section{Ergodicity breaking and charge density waves}\label{sec:CDW}

{The emergence of a new physical regime for sufficiently peaked jump distributions is intimately related to the loss of ergodicity of the Markov chain for Dirac-delta distributions: for 1D systems, Dirac-delta jump distributions of the form $\delta(|\eta|-a)$ only allow jumps of size $\pm a$, thus leading to a discrete points $x_n = x \pm na$, $n=0,1,2...$ in the real line. Therefore, irreducibility is broken and the system is prevented from attaining the equilibrium distribution in the long-time limit. In this section, we show that, for two-peaked jump distributions in 1D, a new regime emerges that solves the above issue. In this regime, the eigenfunctions of the master equation present an oscillatory behaviour, reminiscent of charge density waves {in electronic systems}~\cite{gruner_charge_1985,gruner_dynamics_1988,chen_charge_2016}. {In condensed matter physics, CDWs constitute a collective transport phenomenon that arise from strong electron-phonon interactions in certain low-dimensional materials. Almost a century ago, Peierls suggested that one-dimensional metals are unstable towards the formation of aperiodic lattice distortions associated with spatially periodic modulations of the electronic charge density~\cite{peierls_quantum_1955}---which would be later known as CDWs. The formation of CDWs opens a gap in the Fermi level at low temperatures, which lowers the kinetic energy of the conduction electrons. Due to the similarities between the CDWs found in materials and the explicit form of eigenfunctions reported in this work---i.e. low dimensionality of the system, lattice periodicity and inherent relation to the emergence of a phase transition---we refer to this new regime as the CDW regime.}

{In this section}, we provide a rigorous characterisation of this novel CDW regime {for the Metropolis algorithm in one dimension}, and we corroborate our results by comparing our approximate analytical calculation with the ``exact'' numerical diagonalisation of the master equation\footnote{\label{footnote_lapack}The following numerical procedure has been employed for all the examples presented throughout the paper. To obtain the eigenvalues and eigenfuntions of the master equation, we have employed the FORTRAN 90 subroutine \textit{dsyev.f90} from LAPACK~\cite{anderson_lapack_1999}. To avoid spurious boundary effects, the spatial variable $x$ is restricted to a finite interval $[-L,L]$, with a sufficiently large $L$. Moreover, space is discretised with a mesh of $N_d$ points. For the harmonic potential, we have fixed units by choosing $k_B T=1$ and $\chi=1$, and taken $L=10$ and $N_d=2000$.}

\subsection{Main assumptions behind the emergence of the CDW eigenfunctions from the master equation}\label{subsec:main-assumptions}

{We may} derive the corresponding CDW eigenfunctions directly from the master equation and study their properties. {To do so,} we consider here a two-peak family of jump distributions $\omega(\eta;a,\sigma)$, the maxima of which are at $\eta=\pm a$ with a certain small width, measured by a parameter $\sigma$. The parameters $(a,\sigma)$ define the two characteristic lengths of the jump distribution. Let us introduce the auxiliary distribution
\begin{equation}
    \delta_{\sigma}(\eta-a)\equiv 2\omega(\eta) \theta(\eta).
\end{equation}
By definition, (i) the maximum of $\delta_{\sigma}(\eta)$ is located at $\eta=0$ and (ii) $\delta(\eta)$ identically vanishes for $\eta<-a$. The width of the peak of $\delta_{\sigma}$ is measured by its standard deviation.  Using this definition, we write the jump distribution as follows:
\begin{equation}
    \label{narrow-jump-dis}
    \omega (\eta;a,\sigma) = \frac{1}{2}\left[\delta_{\sigma}(\eta - a) + \delta_{\sigma}(-\eta-a)\right].
\end{equation}
For $\sigma=0$, $\delta_0(\eta)$ corresponds to a Dirac-delta distribution and Eq.~\eqref{narrow-jump-dis} reduces to Eq.~\eqref{eq:omega0-delta-peak}, which---as discussed above---breaks ergodicity and does not allow the system to equilibrate.  Here, we will be mainly interested in the regime $\sigma\ll a$---but non-zero, where the jump distribution is narrowly peaked around $\pm a$, and thus $\omega(0;a,\sigma)=\delta_{\sigma}(-a)\approx 0$.

Now we proceed to analyse the eigenfunctions $\phi_{\nu}(x;a,\sigma)$ of the master equation~\eqref{eq:master-equation}, i.e. the non-trivial solutions of Eq.~\eqref{eq:eigenproblem-2-b-explicit}. It is convenient to introduce the change of variables
\begin{equation}\label{eq:from-phi-to-rho}
    \phi_{\nu}(x;a,\sigma) = \phi_0(x) \rho_{\nu}(x;a,\sigma)\propto e^{-\beta U(x)/2} \rho_{\nu}(x;a,\sigma),
\end{equation}
such that the equilibrium distribution corresponds to a constant, $\rho_0(x;\cancel{a,\sigma})=\text{const.}$ Note that functionals of $\omega$ translate here into a parametric dependence on $(a,\sigma)$, which we omit henceforth to simplify the notation. For the sake of concreteness and simplicity, we present the analysis for confining potentials $U(x)$ that are convex, i.e. $U''(x)>0$, and thus have only one minimum---without loss of generality, we assume that the minimum is at $x=0$. We stress that the convexity condition can be relaxed and our conclusions still hold, but it allows us to simplify the presentation while keeping the main ideas. {
The analytical derivation of the CDW eigenfunctions is rather technical and cumbersome. Hence, we relegate it to Appendix~\ref{app:analytical-derivation}. Below, we outline the two main assumptions employed in the derivation:
\begin{enumerate}
    \item For jumps where the potential energy increases, we assume that the increase is significant. Consequently, most subsequent jumps to the side of the potential will be rejected, while jumps to the opposite side are predominantly accepted. Thus, the overall rejection probability is approximately $1/2$. Mathematically, this applies for sufficiently large values of $a$ and $x$ not too close to the global minimum, such that
    \begin{equation}\label{eq:ansatz-1}
        e^{-\beta\Delta U(x \pm a,x)} \ll 1.
    \end{equation}
    \item We assume that the \textit{charge density wave} (CDW) ansatz,
    \begin{equation}\label{eq:cdw-ansatz}
        \rho_{\nu}(x\pm a) = \rho_{\nu}(x), \ \forall x,
    \end{equation}
    holds. This assumption follows from the fact that, for infinitely sharp jump distributions---i.e., $\sigma = 0$, where ergodicity is broken---the ground state becomes infinitely degenerate. In this case, any periodic solution of the form \eqref{eq:cdw-ansatz} yields the eigenvalue $\lambda_{\nu} = 1$. Thus, our theoretical approach can be regarded as a singular perturbation theory within this subspace.
\end{enumerate}
Based on these assumptions, we expand the master equation from Eq.~\eqref{eq:eigenproblem-2} in powers of $\sigma$, resulting in the following equation for $\rho_{\nu}(x)$:
}

\begin{equation}
    \label{diff-eq}
    \rho_{\nu}''(x) + \frac{4(1-\lambda_{\nu})}{\sigma^2}\rho_{\nu}(x) = 0, \quad \forall x,
\end{equation}
which provides the solutions
\begin{equation}\label{eq:rhonu-cos-sin}
    \rho_{\nu}(x) = c_{\nu}\cos \left( \frac{2\pi \nu}{a}x \right)+d_{\nu}\sin \left( \frac{2\pi \nu}{a}x \right), \quad \lambda_{\nu} = 1 -\nu^2\pi^2 \left( \frac{\sigma}{a} \right)^2.
\end{equation}
Note that these CDW modes only make sense for $\sigma\ne 0$: as stated above, for $\sigma=0$ any $a$-periodic $\rho_{\nu}(x)$ belongs in the subspace corresponding to $\lambda_0=1$, which becomes degenerate due to the breaking of ergodicity. In addition, the CDW modes cease to make sense for very large values of $\nu$: we know that the exact eigenvalues of the master equation cannot become smaller than $-1$, whereas $\lim_{\nu\to\infty}\lambda_{\nu}\to -\infty$~\footnote{A similar issue was found in Ref. \cite{chepelianskii_metropolis_2023} for the Schrödinger eigenvalues.}.

For $\sigma\ne 0$, we have that $\lambda_0=1$ is clearly non-degenerate, since the sine mode identically vanishes for $\nu=0$ and the cosine mode becomes a constant. On the other hand, all the other eigenvalues $\lambda_{\nu}$, with $\nu>0$, seem to be degenerate, since both the sine and the cosine modes are non-zero. Henceforth, we write
\begin{subequations}\label{eigen-CDW}
\begin{equation}\label{eigen-CDW-cos}
    \phi_{\nu}^{\CDW,e}(x) =c_{\nu}^{e}\, \phi_0(x)\cos \left( k_{\nu} x \right)\propto e^{-\beta U(x)/2} \cos \left( k_{\nu} x \right),
\end{equation}
\begin{equation}\label{eigen-CDW-sin}
    \phi_{\nu}^{\CDW,o}(x) =c_{\nu}^{o}\, \phi_0(x)\sin \left( k_{\nu} x \right)\propto e^{-\beta U(x)/2} \sin \left( k_{\nu} x \right), 
\end{equation}
\end{subequations}
with
\begin{equation}
    k_{\nu}=\frac{2\pi\nu}{a}, \quad \lambda_{\nu}=1-\frac{\sigma^2}{4}k_{\nu}^2,
\end{equation}
for the CDW modes that approximate the exact eigenfunctions $\phi_{\nu}(x)$ of the master equation for $\sigma\ll a$. In the above, the indexes $e$ and $o$ refer to the even and odd sectors of the CDW basis, respectively. The normalisation constant $c_{\nu}$ is determined by imposing $\norm{\phi_{\nu}^{\CDW}}=1$, i.e.
\begin{equation}
    \int_{-\infty}^{+\infty} dx \, \left| \phi_{\nu}^{\CDW}(x) \right|^2=1.
\end{equation}

\subsection{Numerical check of the CDW eigenfunctions}\label{sec:numerical-check-CDW}

In order to check the correctness of our theoretical description, we have numerically diagonalised the master equation, i.e.~numerically solved the eigenproblem~\eqref{eq:eigenproblem}, for the harmonic potential with different jump distributions. Again, we have fixed units by taking $k_B T=1$ and $\chi=1$, see~\cite{Note4}. For each of the families of jump distributions, Eqs.~\eqref{eq:parametric-gaussian}--\eqref{eq:a-b-distribution}, we have chosen the parameters so that $\omega(\eta)$ adopts a two-peak structure. 

In Fig.~\ref{fig:cdw-eigen}, the numerical leading eigenfunction $\phi_1(x)$ is compared with the analytical approximation $\phi_1^{\CDW}(x)$ in Eq. \eqref{eigen-CDW}. When it shows oscillatory behaviour, $\phi_1(x)$ is always even: therefore, we have fitted it to a cosine-shaped eigenfunction, 
\begin{equation}
   \phi_1^{\text{fit}}(x)=c(k_1) \phi_0(x) \cos(k_1^* x),
\end{equation} 
where $c(k_1)$ is a normalisation constant such that $\norm{\phi_1^{\text{fit}}}=1$, by minimising the distance between the two functions,
\begin{equation}\label{eq:distance-functions}
    d(\phi_1,\phi_1^{\text{fit}})\equiv \int_{-\infty}^{+\infty} dx \, \left| \phi_1^{\text{fit}}(x)-\phi_1(x)\right|^2
\end{equation}
over $k_1$. The obtained fits are excellent in all cases, with the largest separation taking place near the origin, where the amplitude of the numerical eigenfunction is slightly larger than that of the theoretical curve. This is reasonable, since the equation for the CDW actually holds for $|x|>a/2$, see Appendix~\ref{app:analytical-derivation}. On the one hand, the fitted wave vector $k_1^*=2\pi/a^*$ perfectly matches the theoretical value $k_1=2 \pi/a$ for the Gaussian and box-like jump distributions in Eqs.~\eqref{eq:parametric-gaussian} and \eqref{eq:a-b-distribution}, respectively. On the other hand, $k_1^*\ne 2 \pi/a$ for the algebraic jump distribution in Eq.~\eqref{eq:alpha-a-distribution}. We recall that we described $\delta_{\sigma}(\eta-a)$---which is basically the quantity shown in the lower right panel of Fig.~\ref{fig:cdw-eigen}---as a peaked distribution around $\eta=a$ with a width proportional to its variance. Although this is a fair description of Gaussian and box-like distributions, it is not so for algebraic jump distributions, which are not symmetric around the point at which the probability accumulates as $\sigma\to 0^+$. 
However, a good theoretical prediction for $k_1^*$ is always given by $k_{\text{th}}^*=2\pi/a_{\text{th}}^*$, with $a_{\text{th}}^*$ being the point at which the cumulative distribution equals $1/2$, i.e. $\int_{0}^{a_{\text{th}}^*}d\eta\, \delta_{\sigma}(\eta-a)=1/2$. For the parameters employed in the figure, $a_{\text{th}}^*=1.58$ while the fitted value is $a^*=1.56$---the difference is approximately $1\%$.
Note that $a^*_{\text{th}}=a$ for both the Gaussian and box-like distributions in Eqs.~\eqref{eq:parametric-gaussian} and \eqref{eq:a-b-distribution}.
\begin{figure*}
  \centering 
 {\includegraphics[width=3.1in]{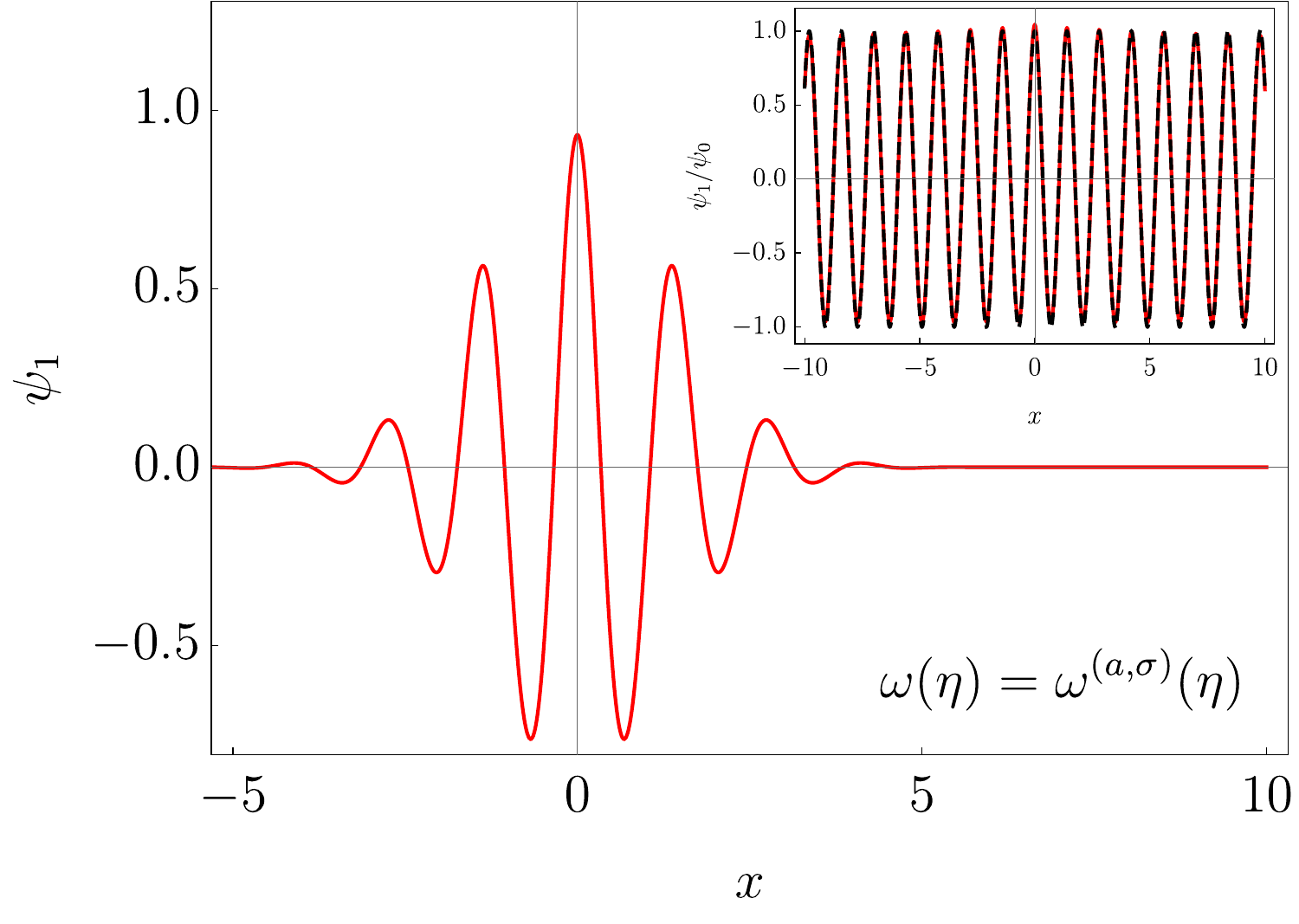} 
 \includegraphics[width=3.1in]{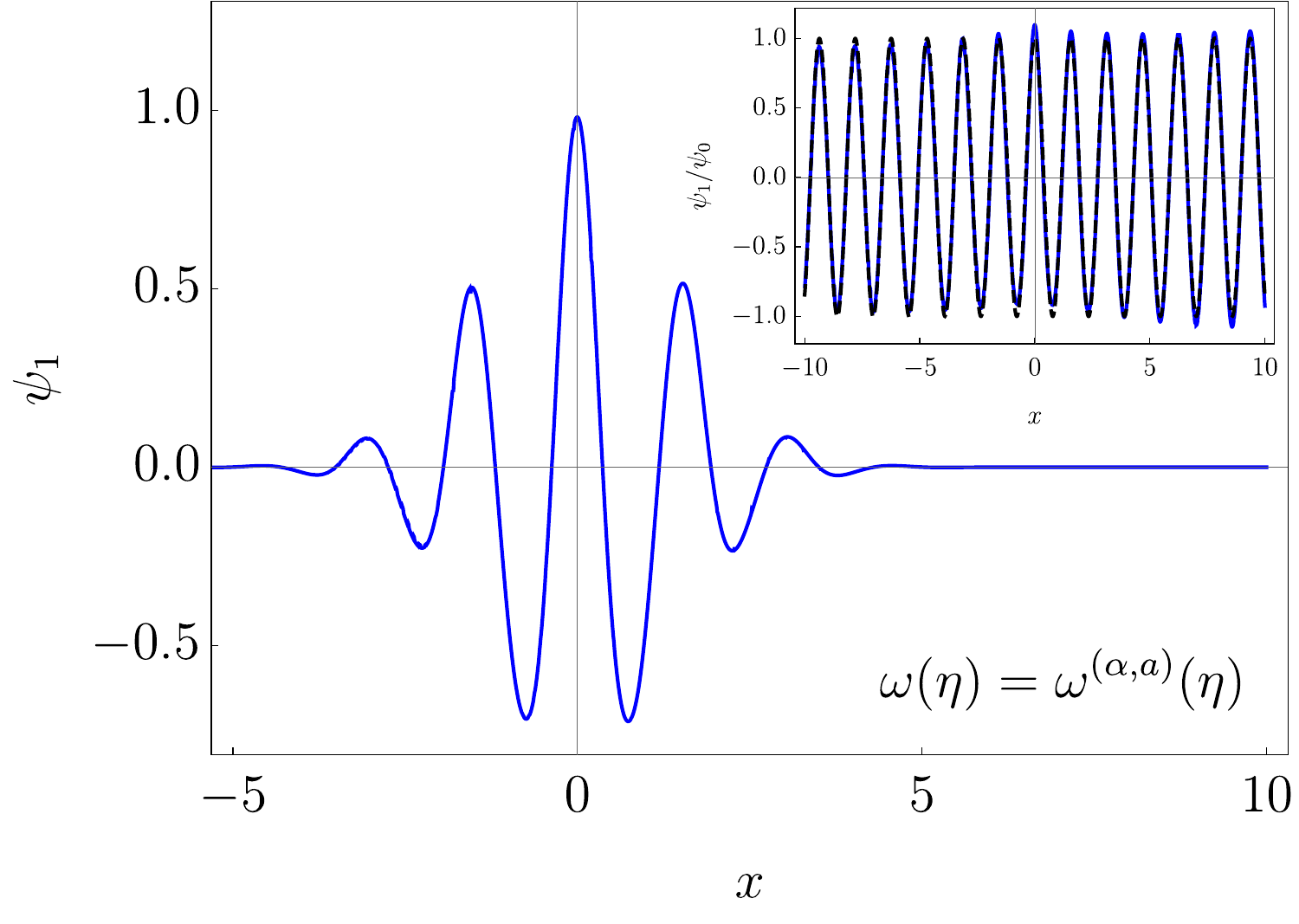}\\
 \includegraphics[width=3.1in]{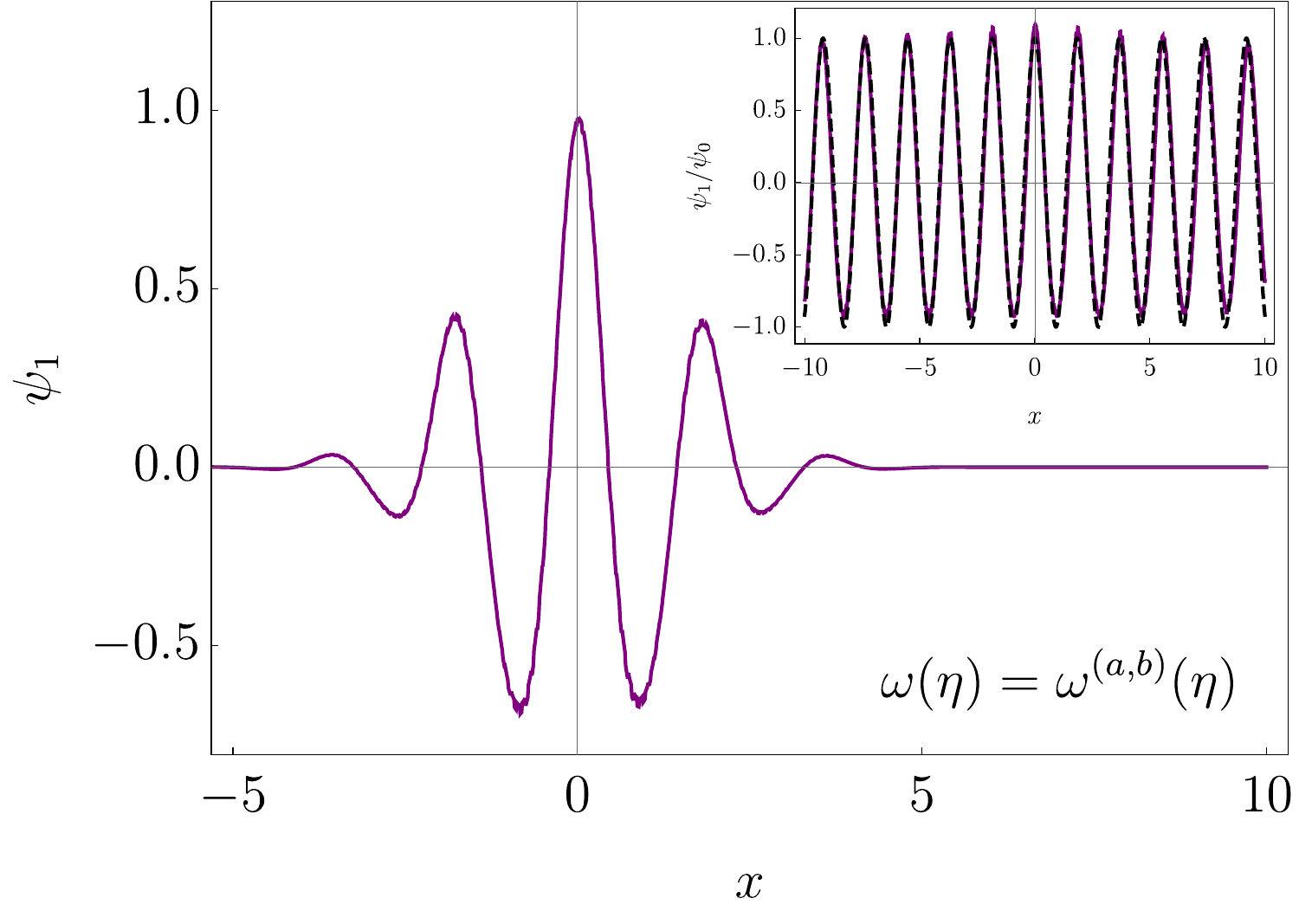} 
 \includegraphics[width=3.0in]{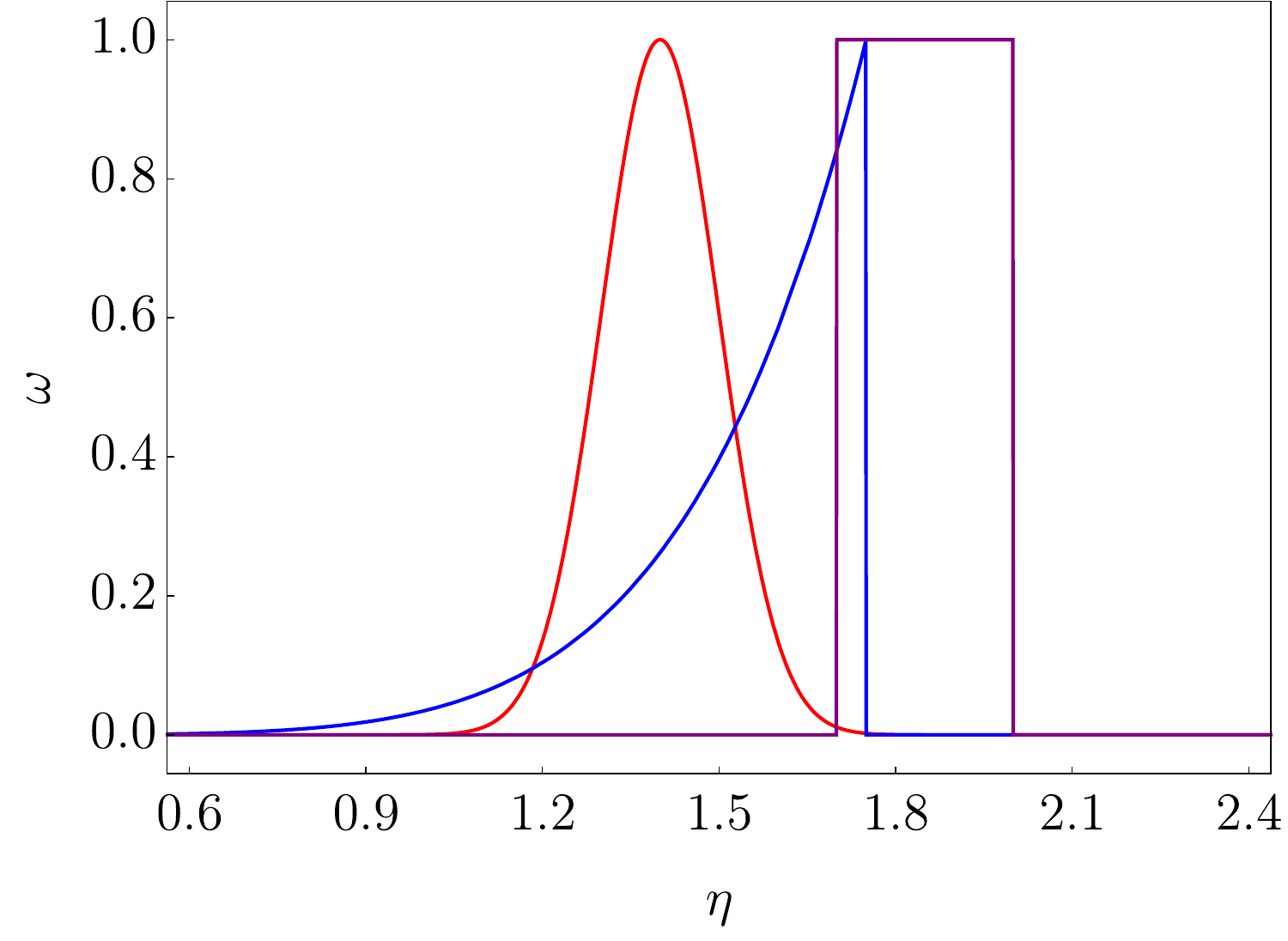}}
  \caption{Leading eigenfunction for the Metropolis algorithm. It has been obtained via numerical diagonalisation of the master equation, for: (upper left) $\omega^{(a,\sigma)}(\eta)$, with $(a,\sigma) = (1.4,0.1)$, (upper right) $\omega^{(a,\alpha)}(\eta)$, with $(a,\alpha) = (1.75,6)$, and (lower left) $\omega^{(a.b)}(\eta)$ with $(a,b) = (1.85,0.15)$. The insets present the ratio between the leading eigenfunction and the equilibrium distribution $\phi_0(x)$. The dashed curves correspond to the best fit of such ratios to the function $\cos(k_1^* x)$, with $k_1^*=2\pi/a^*$, with (upper left) $a^* = 1.40$, (upper right), $a^* = 1.56$, and (lower left) $a^* = 1.85$,  respectively. The corresponding jump distributions---only for $\eta>0$, recall that they are even functions of $\eta$---are shown in the lower right panel: $\omega^{(a,\sigma)}(\eta)$ in red, $\omega^{(a,\alpha)}(\eta)$ in blue, and $\omega^{(a,b)}(\eta)$ in purple.
  }
  \label{fig:cdw-eigen}
\end{figure*}

It is interesting that the leading eigenfunction always seems to be even when the CDW oscillatory behaviour emerges. This fact hints at the terms neglected in our derivation of the CDW modes breaking the degeneracy of the sine and cosine eigenfunctions. A more refined approach would be necessary to analytically explain the even parity of the leading eigenfunction.

\section{Phase-diagrams for the diffusion-rejection-cdw picture}\label{sec:phase-diag}

In this section, we apply our findings in the previous section to investigate the optimisation of the convergence rate for one family of bi-parametric jump distributions, and explore the different phases or regimes that emerge when varying the system parameters. Specifically, we apply our general results to the specific case of harmonic confinement, with the algebraic family of jump distributions of Eq.~\eqref{eq:alpha-a-distribution}. 

\subsection{Order parameters}\label{subsec:order-parameters}

Our goal is to draw the phase diagram of the system in the $(a,\alpha)$ plane of system parameters. In that way, we will elucidate the interplay between the diffusion, rejection, and CDW phases. The diffusive phase corresponds to the region in the $(a,\alpha)$ plane where the leading eigenvalue $\Lambda$ is given by the diffusive modes---analogously, we speak of the rejection and CDW phases. As one moves in the $(a,\alpha)$ plane, the jump distribution changes shape. For distributions that allow only small jumps, the diffusive phase is expected; for distributions that allow for large jumps, the rejection phase is expected; for two-peak distributions, the CDW phase is expected. One key finding of our analysis is that the point $(a^*,\alpha^*)$ at which the leading eigenvalue $\Lambda$ takes its minimum value in the plane is the triple point of phase coexistence.

In order to characterise the different phases, i.e. diffusion-dominated, rejection-dominated, and CDW-dominated, we introduce the following two order parameters, the fidelity $\calF$ and the inverse participation ratio IPR:
\begin{itemize}
    \item \textbf{Fidelity}: Similarly to that employed in the context of coherent states in quantum mechanics, we define the fidelity for the problem of our concern as
    \begin{equation}
    \label{fidelity}
        \calF \equiv \left|\braket{\phi_{1}}{\phi_1^{\CDW}}\right|^2 = \underset{k}{\max}\left|\int_{-\infty}^{+\infty}dx \ c_1 \phi_0(x) \cos (kx)  \phi_{1}(x)\right|^2,
    \end{equation}
    which measures the affinity between the actual leading eigenfunction $\phi_{1}(x)$ of the Metropolis algorithm with the CDW ansatz $\phi_1^{\CDW}$ from Eq.~\eqref{eigen-CDW} with the corresponding optimal wave number $k^*$. Let us note that the fidelity is closely related to the distance about the involved functions, $d=2(1-\calF)$---compare Eq.~\eqref{fidelity} with Eq.~\eqref{eq:distance-functions}.
    
    The case $\calF=0$ corresponds to null affinity, while $\calF=1$ results in a perfect match. For $k^* \rightarrow 0$, the fidelity tends to zero, since the actual leading eigenfunction $\phi_{1}(x)$ is orthogonal to the equilibrium solution. 
    \item \textbf{Inverse Participation Ratio}: This quantity, introduced in Ref.~\cite{chepelianskii_metropolis_2023}, measures the emergence of the localisation transition in the eigenfunctions of the master equation. In our numerical diagonalisation of the master equation, space is discretised into a lattice with $N_d$ sites. Within this discretisation scheme, the eigenfunction $\phi_{\nu}(x)$ simplifies into a set of $N_d$ values $\phi_{\nu}(i)$, $i=1,\ldots,N_d$. The IPR for the leading eigenfunction is 
    \begin{equation}\label{eq:IPR}
        \IPR \equiv \sum_{i=1}^{N_d}|\phi_{1}(i)|^4 \bigg/ \left(\sum_{i=1}^{N_d}|\phi_{1}(i)|^2\right)^2.
    \end{equation}
    If $\phi_{1}(x)$ is completely delocalised, then $\IPR \rightarrow 0$ for large $N_d$, implying that it has a well-defined continuum limit, with no delta peaks. If part of the eigenfunction gets localised, then $\IPR = O(1)$.
\end{itemize}

\subsection{Numerical results on the phase behaviour}\label{subsec:num-results-phase-behaviour}
In Figs.~\ref{fig9} and \ref{fig8}, the heat maps for the fidelity $\calF$, the IPR and {the next-to-largest eigenvalue} $\Lambda$ are shown on the $(a,\alpha)$ plane for the bi-parametric family of algebraic jump distributions in Eq.~\eqref{eq:alpha-a-distribution}. These diagrams have been obtained by numerically diagonalising the master equation. We recall that the parameter $a$ gives the typical size of the attempted jumps while the parameter $\alpha$ controls the width of the distribution, which becomes a delta peak at $\eta=a$ in the limit as $\alpha\to\infty$.

In Fig.~\ref{fig9}, we show the heat map of the order parameters, the fidelity $\calF$ (left) and the IPR (right). 

\begin{itemize}
    \item \textbf{Behaviour of the fidelity}: For low values of $\alpha$, the fidelity is always zero. The jump distribution does not have the double-peak structure, and then the CDW modes do not emerge. Instead, for $\alpha>\alpha_c=1.22$, $\calF$ abruptly varies from close-to-zero values to close-to-unity values, when increasing $a$.  For $a<a_1(\alpha)$, the Schrödinger modes dominate and we are in the diffusive regime, for which $\calF\approx 0$, while for $a > a_1(\alpha)$, the CDW modes emerge and dominate, and thus $\calF\approx 1$. The fidelity order parameter is discontinuous at the line $a=a_1(\alpha)$, which is thus a first-order transition line. As we continue increasing $a$, the fidelity begins to smoothly decrease back towards zero around a certain value $a_2(\alpha)$ that is less clearly defined. This is related to the change from the CDW regime to the rejection regime---as we will see more clearly when considering the IPR. 
    
    \item \textbf{Behaviour of the IPR}: For $\alpha<\alpha_c$, for which the CDW regime does not emerge, the localisation transition takes place along the line $a=a_3(\alpha)$. Thereat, the IPR abruptly changes from almost null values to $O(1)$ ones, and $a=a_3(\alpha)$ is another first-order transition line---as already reported in Ref.~\cite{chepelianskii_metropolis_2023}. For $\alpha>\alpha_c$, the IPR vanishes both in the regions where the diffusive modes and the CDW modes dominate. As $a$ increases and approaches the region where the fidelity smoothly decreases back from unity to zero, the IPR in turn smoothly increases from zero to unity---and retains non-zero, $O(1)$, values when $a$ is further increased. The transition line is not as clear-cut as $a_1(\alpha)$ and $a_3(\alpha)$, since both $\calF$ and the IPR vary smoothly. As a first approach, we could define the second-order transition line $a_2(\alpha)$ as the locus at which $\calF=$IPR. The three so-defined transition lines cross at the tricritical point $(a_c,\alpha_c) = (2.08,1.22)$.
\end{itemize}

\begin{figure}
  \centering
  \includegraphics[width=3.3in]{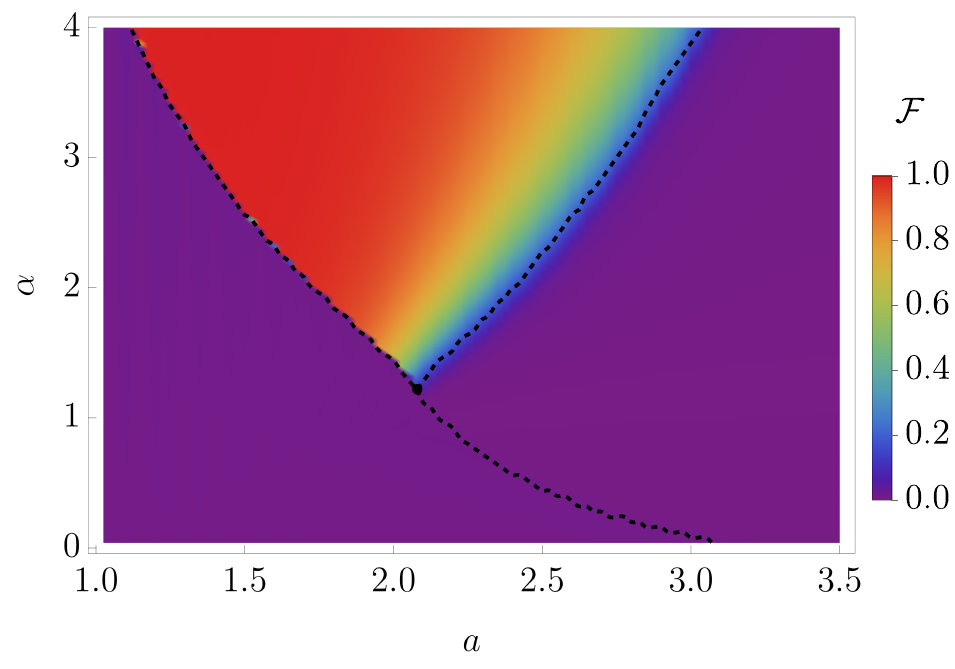}
  \includegraphics[width=3.3in]{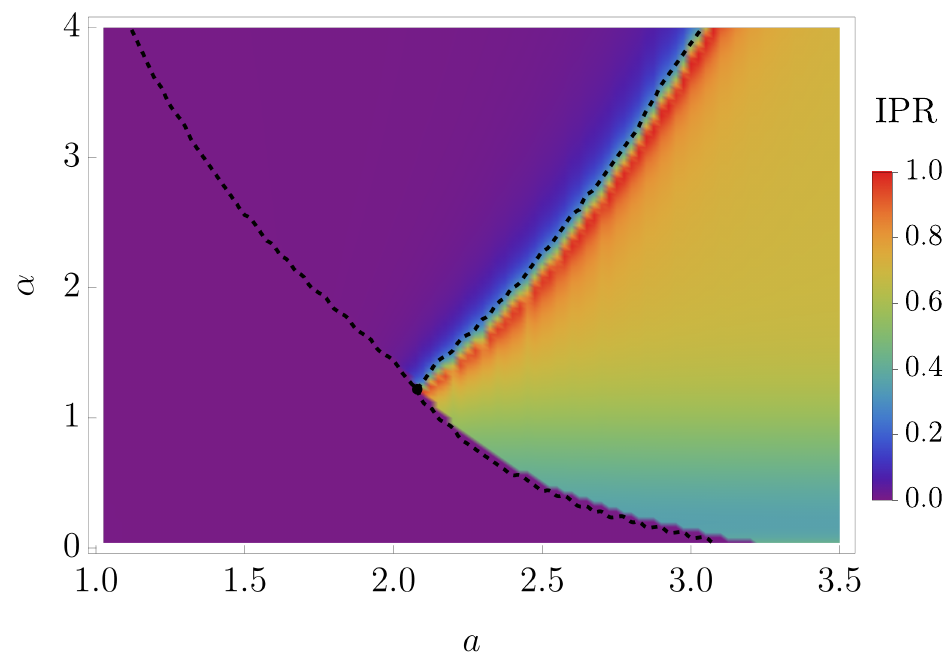}
  \caption{Heat maps of the fidelity $\calF$ (left) and the IPR (right). Both are plotted in the $(a,\alpha)$ plane of parameters for the algebraic jump distribution in Eq.~\eqref{eq:alpha-a-distribution}. Both order parameters have been evaluated by numerically obtaining the leading eigenfunction of the master equation. The transition lines between the different phases are plotted with dashed lines and intersect at the tricritical point $(a_c,\alpha_c) = (2.08,1.22)$, which is represented by a black dot. These lines correspond to (i) $a_1(\alpha)$ for $\alpha > \alpha_c$ and $a<a_c$, (ii) $a_2(\alpha)$ for $\alpha > \alpha_c$  and $a>a_c$, and (iii) $a_3(\alpha)$ for $\alpha < \alpha_c$ and $a>a_c$, respectively, as defined in the text.
  }
  \label{fig9}
\end{figure}

Figure~\ref{fig8} shows the corresponding heat map of the {next-to-largest eigenvalue} $\Lambda$, {obtained by numerically diagonalising the master equation. We have checked that, up to our numerical precision, $\Lambda$ attains its global minimum, $\Lambda_{\min}\simeq 0.6170$, at the tricritical point.} As is neatly seen, there exists a valley-like region for $\alpha \leq 2$ in which $\Lambda$ has values that are very close to the minimum. The valley comprises points close to the transition line (i) $a_1(\alpha)$ between the diffusive and CDW phases, for $\alpha_c< \alpha<2$, and (ii) $a_3(\alpha)$ between the diffusive and rejection phases, for $\alpha<\alpha_c$. 
\begin{figure}
  \centering
  \includegraphics[width=4.5in]{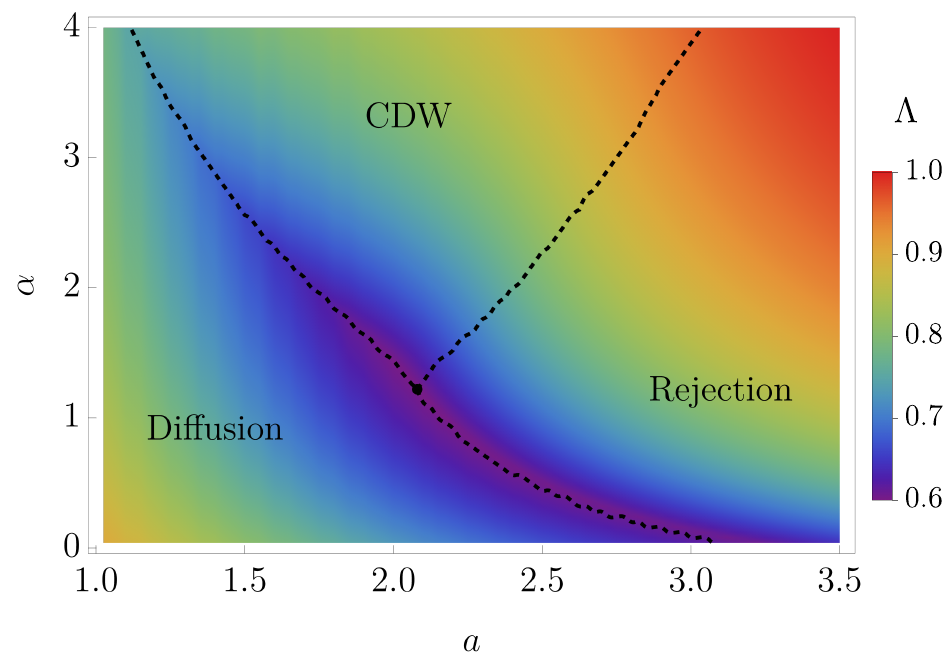}
  \caption{
  Heat map of the {next-to-largest eigenvalue} $\Lambda$ in the $(a,\alpha)$ plane. As in Fig.~\ref{fig9}, the dashed curves correspond to the transition lines between phases. The black point indicates the coordinates of the global minimum, $(a_c,\alpha_c) = (2.08,1.22)$, which coincides with the tricritical point at which the transition lines intersect. 
  }
  \label{fig8}
\end{figure}


{ Figure~\ref{fig_eigenvalues} gives us further information about the phase transitions described above. {Specifically, we plot the fidelity $\mathcal{F}$ and the IPR for the $(a,\alpha)$-family of jump-distributions from Eq.~\eqref{eq:alpha-a-distribution}, by varying $a$ for fixed $\alpha = 3$---similarly to right panel of Fig.~\ref{fig_eigenvalue-crossing}}. {For small values of $a$, we are in the diffusive regime, for which the fidelity and the IPR are close to zero. At the transition point $a_1(\alpha = 3) = 1.37$ between the diffusive and CDW regimes, eigenvalue crossing takes place---as shown in Fig.~\ref{fig_eigenvalue-crossing},} such that the dominant eigenfunction $\phi_1(x)$ changes parity: from odd, for $a<a_1$, to even, for $a>a_1$. This is consistent with the behaviour of the dominant eigenfunctions in the respective phases: in the diffusion regime, the leading eigenfunction $\phi_1(x)$ is provided by the odd sector, whereas, in the CDW regime, the leading eigenfunctions is even---as discussed in Sec.~\ref{sec:numerical-check-CDW}. Upon further increase in $a$, the eigenvalues smoothly tend to the rejection curve, marking the onset of the rejection phase. {The abrupt changes in the behaviour among the diffusive, CDW and rejection regimes reflect the emergence of different underlying phase transitions in the relaxation dynamics of the Metropolis algorithm.} First, the transition between the diffusive and the CDW regimes constitutes a first-order phase transition, characterised by {a} discontinuous change of {the }fidelity, which is due to the fact that the takeover of the leading relaxation mode involves a change of symmetry in the leading eigenfunction. Second, the transition from the CDW regime to the rejection regime constitutes a higher-order---at least second-order---phase transition, characterised by the smooth changes of both $\calF$ and IPR, which is due to the also smooth variation of the leading eigenvalue between the two regimes.}
\begin{figure}
  \centering
  \includegraphics[width=4.5in]{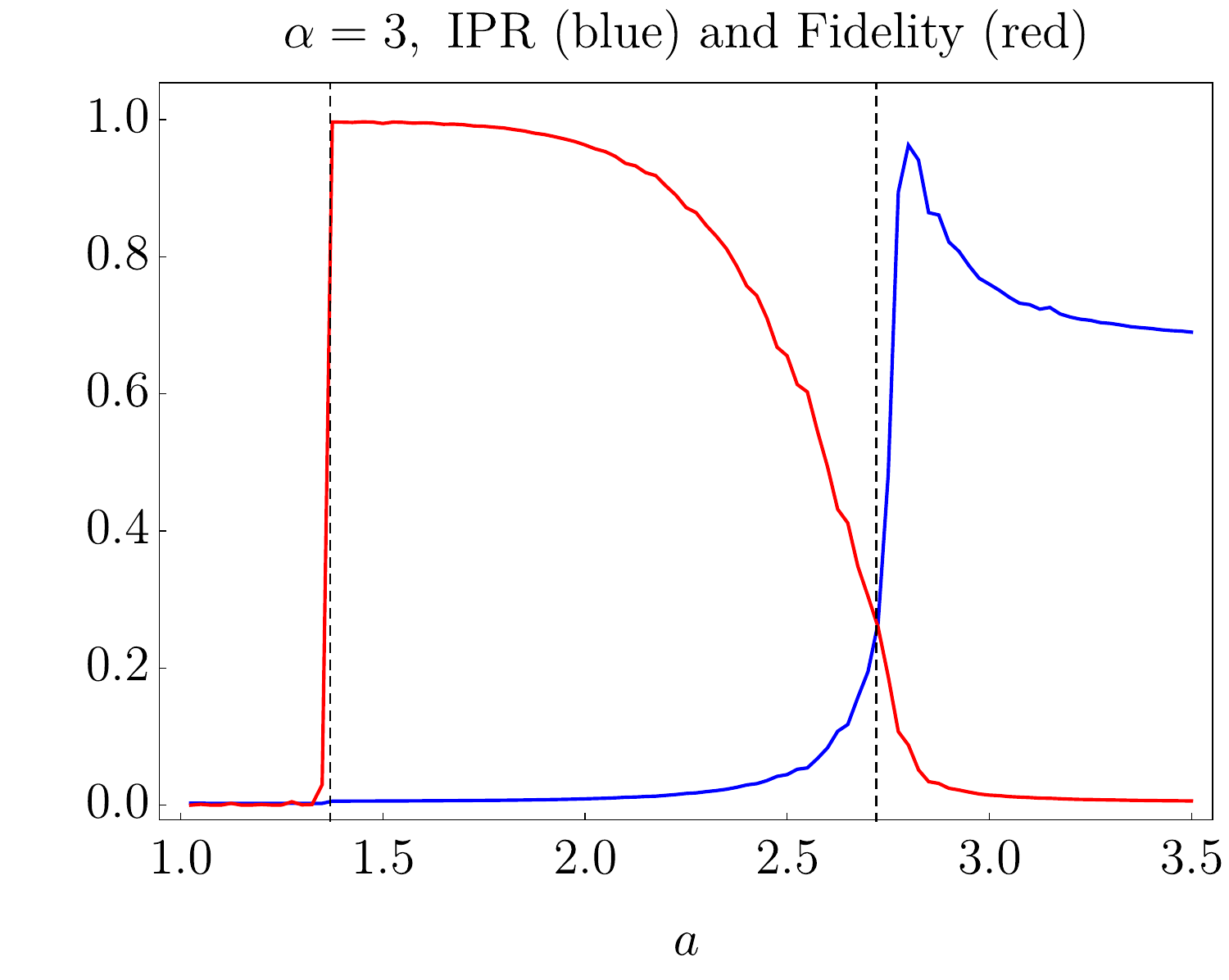}
  \caption{{Fidelity $\calF$ and IPR as functions} of $a$, for a fixed value of $\alpha=3>\alpha_c${---similar to the right panel of Fig.~\ref{fig_eigenvalue-crossing}. Specifically}, we have employed the jump distribution $\omega^{(a,\alpha)}(\eta)$ in Eq.~\eqref{eq:alpha-a-distribution}. The dashed, vertical lines delimit the three different phases explored in this work: for $a<a_1(\alpha = 3)=1.37$, the Schrödinger diffusion-like regime, for $a_1(\alpha = 3)=1.37<a<a_2(\alpha = 3)=2.72$, upon eigenvalue crossing, the CDW regime, and for $a>a_2(\alpha = 3)=2.72$, the rejection regime.
  }
  \label{fig_eigenvalues}
\end{figure}

\section{Estimating the optimal convergence rate}\label{sec: optimality}

{As we may infer from the approaches followed in the previous sections, analytically determining both $\lambda_{1}[\omega]$ and its corresponding eigenfunction $\phi_{1}[\omega](x)$ from the integral equation~\eqref{eq:eigenproblem} is generally not possible. Needless to say it is that the same applies to determining the optimal jump distribution $\omega_{\text{opt}}(\eta)$ that maximises the convergence rate. In the remaining sections, we employ a variational approach in order to estimate the convergence rate and the optimal jump distribution for different families of jump distributions. The advantages of this approach are twofold: (i) it is computationally less expensive than the numerical diagonalisation method employed to solve the master equation, and (ii) as we will show, it allows us to further corroborate the diffusion-rejection-CDW picture.}

{Let us consider a certain known basis of orthogonal functions $\psi_n(x)$, and expand the function $\phi(x)$ in Eq.~\eqref{functional-true} in this basis. Truncating the expansion after $N_s$ states, which must be perpendicular to the ground state, we get a lower bound for $\Lambda[\omega]$, i.e.
\begin{equation}
\label{lower-bound}
    \Lambda[\omega] \geq \tilde{\Lambda}[\omega]\equiv \underset{c_1, c_2,\ldots, c_{N_s}}{\max} \ \int dx \int dx' \ \phi(x)\Kop\phi(x'), \quad \phi(x) = \sum_{n=1}^{N_s} c_n \psi_n(x).
\end{equation}
Let us remark that the above inequality saturates only in the limit $N_{s}\rightarrow +\infty$. Therein, we have the complete expansion of $\phi(x)$, and therefore the equality is attained regardless of our basis choice. However, certain choices of basis, i.e. certain choices of the functions $\psi_n(x)$---which we refer to as physical choices---will get us closer to the true next-to-largest eigenvalue than others, for a not so large number of retained eigenmodes $N_s$.

Now, in order to calculate our estimate $\tilde{\Lambda}[\omega]$ for the next-to-largest eigenvalue, we must find the optimal values of the coefficients $\{ c_n \}_{n=1}^{N_s}$ that maximise the integral in Eq.~\eqref{lower-bound}. Such problem is equivalent to finding the largest eigenvalue of the reduced $N_s \times N_s$ matrix $\tilde{K}[\omega]$ with elements
\begin{equation}
\label{mat}
    \tilde{K}_{nm}[\omega] = \int dx \int dx' \ \psi_n(x)\Kop \psi_m(x'),
\end{equation}
corresponding to the truncation $1\leq n,m \leq N_s$. 
It is worth emphasising that for symmetric confining potentials, the eigenbasis of our problem may be split into even and odd eigenfunctions. The integral kernel $\Kop$ inherits these symmetries, implying that the matrix elements $\tilde{K}_{nm}[\omega]$ are zero for those values of $n$ and $m$ for which the corresponding eigenfunctions have different parity. Thus, we may divide the basis into two bases: one containing the even eigenfunctions, and the other one containing the odd eigenfunctions. In this way, the matrix $\tilde{K}_{nm}$ can be written as a direct sum of even and odd sectors.

Using different bases, we can ensure that our lower bound is sufficiently close to the true {next-to-largest eigenvalue} for every possible jump distribution $\omega(\eta)$. In fact, using $M$ different bases, we may define the functional
\begin{equation}
\label{general-lower-bound}
    {\tilde{\Lambda}}_{M}[\omega] \equiv \max \left[{\tilde{\Lambda}}^{(1)}[\omega],{\tilde{\Lambda}}^{(2)}[\omega],..., {\tilde{\Lambda}}^{(M)}[\omega] \right]
\end{equation}
to estimate the next-to-largest eigenvalue, taking a sufficiently high number of retained eigenfunctions $N_s$ for each basis. In what follows, we mainly use the estimator above with different values of $M$, {typically with $100\le N_s\le 300$} retained eigenfunctions per basis. 

In order to approximately solve the min-max problem presented above, we thus resort to a two-step procedure. First, for each considered jump distribution $\omega(\eta)$, we use the functional~\eqref{general-lower-bound} to find an accurate estimate for the leading eigenvalue $\lambda_1[\omega]$, i.e. the maximum ${\tilde{\Lambda}}_M[\omega]$ of its lower bounds ${\tilde{\Lambda}}^{(i)}[\omega]$, $i=1,\ldots,M$, for the $M$ bases considered. Second, we vary the jump distribution to find the minimum value of such estimates ${\tilde{\Lambda}}_M^*$, which is attained for a specific jump distribution $\omega_M^*$, i.e. ${\tilde{\Lambda}}_M^*\equiv {\tilde{\Lambda}}_M[\omega_M^*]$~\footnote{To numerically estimate $\Lambda$, as defined in Eq.~\eqref{simp1-lower-bound}, we have employed the open source nonlinear optimisation libraries from \textit{NLopt} \cite{johnson_nlopt_2007} with $N_s=100$ modes for each basis.}. One could use bases without a clear physical meaning and the lower bound obtained in this way might still be quite useful for numerical purposes, but this ``blind'' procedure would not provide meaningful insights into the behaviour of the eigenfunctions of the master equation and the physics underlying the optimal value for the convergence rate---we recall that the convergence rate is determined by the next-to-largest eigenvalue, as given by Eqs.~\eqref{eq:gamma-def} and \eqref{lambda*-def}.}

\subsection{Estimators for the next-to-largest eigenvalue}\label{sec:estimator}

{Following the Schrödinger basis obtained for sufficiently narrow jump distributions around $\eta = 0$---briefly reviewed at the beginning of Sec.~\ref{sec:localisation-and-collapse}, the next-to-largest eigenvalue $\Lambda$ was estimated in Ref.~\cite{chepelianskii_metropolis_2023} by using the 3-bases functional
\begin{equation}
    \label{simp1-lower-bound}
    {\tilde{\Lambda}}_{3}[\omega] \equiv \max \left[ {\tilde{\Lambda}}^{(\text{Sch.})}_{\text{o}}[\omega], {\tilde{\Lambda}}^{(\text{Sch.})}_{\text{e}}[\omega],\underset{x}{\max}\ R[\omega](x)\right],
\end{equation}
where ${\tilde{\Lambda}}^{(\text{Sch.})}_{\text{o}}[\omega]$ and ${\tilde{\Lambda}}^{(\text{Sch.})}_{\text{e}}[\omega]$ are the estimates for $\Lambda$ obtained from the odd and even sectors of the Schr\"odinger basis, respectively. Estimate~\eqref{simp1-lower-bound} is based on the assumption that the convergence rate is obtained as a balance between the diffusion modes of the Schrödinger equation and the maximum value of the rejection rate. The validity of such an argument was checked for several choices of the confining potential $U(x)$ and different jump distributions $\omega(\eta)$, 
even for a few number $N_s$ of retained eigenmodes.  

Given the results previously presented concerning the characterisation of the CDW regime, we} build another estimator for the next-to-largest eigenvalue by adding the CDW basis, as given by Eq.~\eqref{eigen-CDW}, thereto:
\begin{equation}
    \label{eq:new-functional}
    {\tilde{\Lambda}}_{4}[\omega] \equiv \max \left[ {\tilde{\Lambda}}^{(\text{Sch.})}_{\text{o}}[\omega],{\tilde{\Lambda}}^{(\text{Sch.})}_{\text{e}}[\omega],{\tilde{\Lambda}}^{(\CDW)}[\omega],\underset{x}{\max}\ R(x;\omega)\right],
\end{equation}
where ${\tilde{\Lambda}}^{(\CDW)}[\omega]$ is the estimate given by the CDW basis for the next-to-largest eigenvalue.

Let us also consider the localised Gaussian modes 
\begin{equation}
        \phi_n(x) = \mathcal{N}_n \left[e^{-\frac{(x-a_n)^2}{2\sigma^2}} + \epsilon e^{-\frac{(x+a_n)^2}{2\sigma^2}}  \right], \quad a_n = \frac{L}{N_s}\left(n + \frac{1}{2}\right), \quad \sigma = \frac{L}{N_s}, \quad n=0,1,...,N_s-1,
\end{equation}
where $\mathcal{N}_n$ is a normalisation constant. The parameter $\epsilon$ is equal to either $+1$ or $-1$ and sets the symmetry of the eigenfunction. Therefore, this basis can also be split into even and odd sectors.
By adding the even and odd bases of localised Gaussian modes to ${\tilde{\Lambda}}_4$, we build the more refined estimator with $M=6$ bases:
\begin{equation}
    {\tilde{\Lambda}}_6[\omega] \equiv \max \left[ {\tilde{\Lambda}}^{(\text{Sch.})}_{\text{o}}[\omega], {\tilde{\Lambda}}^{(\text{Sch.})}_{\text{e}}[\omega], {\tilde{\Lambda}}^{(\text{Gau.})}_{\text{e}}[\omega],{\tilde{\Lambda}}^{(\text{Gau.})}_{\text{o}}[\omega],{\tilde{\Lambda}}^{(\CDW)}[\omega],\underset{x}{\max}\ R(x; \omega)\right],
\end{equation}
where ${\tilde{\Lambda}}^{(\text{Gau.})}_{\text{e}}[\omega]$ and ${\tilde{\Lambda}}^{(\text{Gau.})}_{\text{o}}[\omega]$ are the estimates for the next-to-largest eigenvalue provided by the even and odd sectors, respectively, of the localised Gaussian modes basis.

We recall the min-max problem that we solve for estimating {the next-to-largest eigenvalue $\Lambda$.} For each jump distribution $\omega$, the above estimators have a definite value---the maximum over the $M$ bases considered in each estimator, which we denote by ${\tilde{\Lambda}}_M[\omega]$. Afterwards, by varying the jump distribution, we find out the optimal jump distribution that minimises ${\tilde{\Lambda}}_M[\omega]$: this is what we denote by $\omega_M^*$, which in turn provides the estimate ${\tilde{\Lambda}}_M^*\equiv {\tilde{\Lambda}}_M[\omega_M^*]$ for the considered set of bases. Note that we always have ${\tilde{\Lambda}}_6^*\geq {\tilde{\Lambda}}_4^*\geq {\tilde{\Lambda}}_3^*$; the functional ${\tilde{\Lambda}}_6[\omega]$ provides the most accurate lower bound for the next-to-largest eigenvalue.

{In the following sections, the values of $\tilde{\Lambda}_M[\omega]$ and $\tilde{\Lambda}_M^*$ are given to $10^{-4}$ precision. This is a challenging precision, but one that can be obtained by increasing the number of retained modes $N_s$ for each basis. With the typical size employed throughout this work, $100\le N_s\le 300$, we have checked that further increasing $N_s$ do not alter the reported values for our best estimate $\tilde{\Lambda}_6^*$ of the next-to-largest eigenvalue---recall that any basis would perfectly give $\Lambda$ in the limit as $N_s\to\infty$.}

\subsection{Corroborating the physical picture}\label{sec:fix-collapse-ins}

To start with, {let us consider the bi-parametric $(a,\sigma)$ family of Gaussian jump distributions defined in Eq.~\eqref{eq:parametric-gaussian} for the Metropolis algorithm. That is, we look for the optimal values of the parameters $(a^*,\sigma^*)$ that minimise the next-to-largest eigenvalue $\Lambda$, following our variational approach. In the numerical scheme for the optimisation, a lower cut-off in the width $\sigma$ must be imposed, i.e. $\sigma\ge \sigma_{\cutoff}$, since very small values of $\sigma$ lead to numerical difficulties~\footnote{Since the numerics involve a spatial discretisation, $\sigma$ must be larger than the lattice spacing in this discretisation.}. If, when decreasing $\sigma_{\cutoff}$, $\sigma^*$ consistently reached the cut-off value, this would hint at an incomplete convergence of $\sigma$ due to the cut-off, i.e. it would hint at actually having $\sigma^*\to 0$. We refer to this problematic scenario as the \textit{collapse instability}, since the optimal jump distribution seems to converge to a double Dirac-delta peak, which breaks ergodicity as {discussed in the preceding sections}.} 

In Fig.~\ref{fig:instabilities-fixed}, we display the optimal jump distributions obtained via the functionals ${\tilde{\Lambda}}_3$, ${\tilde{\Lambda}}_4$ and ${\tilde{\Lambda}}_6$ for the next-to-largest eigenvalue---note that the latter two functionals include the CDW basis. The cutoff $\sigma_{\cutoff}=0.1$ has been employed when finding the optimal parameters $(a,\sigma)$ for each estimator. {The failure of the diffusion-rejection picture is illustrated by the fact that $\omega_3^*(\eta)$ reaches the numerical cutoff $\sigma_3^* = \sigma_{\text{cutoff}}$, thus tending towards a Dirac delta-like distribution.} It is neatly seen that the variances of $\omega_4^*(\eta)$ and $\omega_6^*(\eta)$ no longer reach the cutoff, so the collapse instability is fixed {by including the CDW basis}. Moreover, $\omega_4^*$ and $\omega_6^*$ are close and quite different from the Dirac-like distribution $\omega_3^*$. On the one hand, all the functionals give estimates for the next-to-largest eigenvalue that are not so different, ${\tilde{\Lambda}}_3^*=0.6114$, ${\tilde{\Lambda}}_4^*=0.6138$, and ${\tilde{\Lambda}}_6^*=0.6155$. On the other hand, the incomplete convergence of $\omega_3^*$ can further be verified by inserting it into ${\tilde{\Lambda}}_4[\omega]$ and ${\tilde{\Lambda}}_6[\omega]$. This gives ${\tilde{\Lambda}}_4[\omega_3^*]=0.9239$ and ${\tilde{\Lambda}}_6[\omega_3^*]=0.9521$, which have a relative error of more than $50\%$ with respect to our best estimate of the  next-to-largest eigenvalue ${\tilde{\Lambda}}_6^*$. If we do a similar check with $\omega_4^*$, by evaluating ${\tilde{\Lambda}}_6[\omega_4^*]=0.6239$, the relative error is reduced to $1.36\%$, thus clearly showing the improvement of the convergence obtained with the incorporation of the CDW basis. 
\begin{figure}
  \centering
  \includegraphics[width=3.3in]{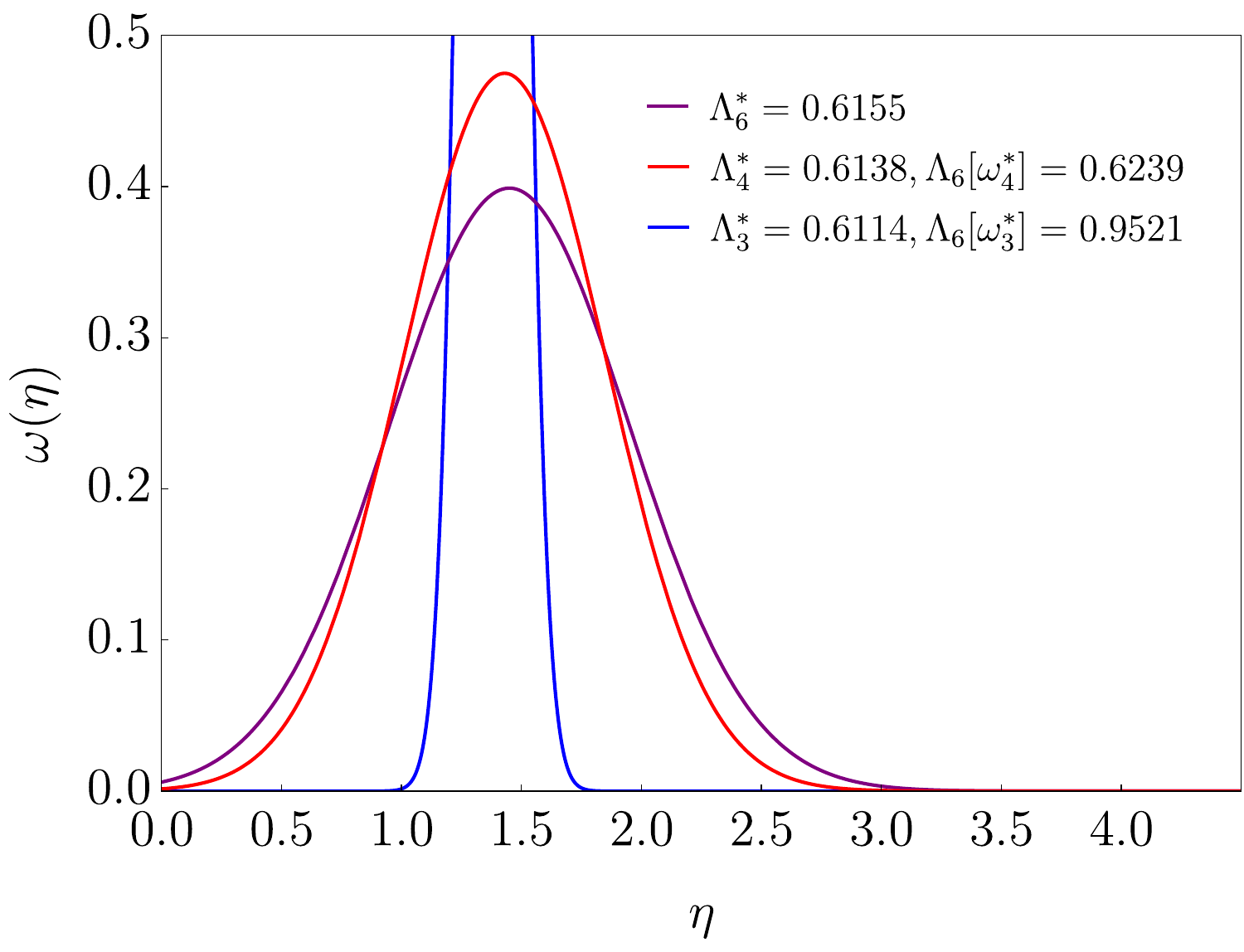}
  \caption{Optimal jump distributions for the Gaussian family in Eq.~\eqref{eq:parametric-gaussian}. These optimal jump distributions have been obtained by employing the three functionals ${\tilde{\Lambda}}_3[\omega]$, ${\tilde{\Lambda}}_4[\omega]$, and ${\tilde{\Lambda}}_6[\omega]$ for the {next-to-largest eigenvalue}. Specifically, we show $\omega_6^*$ (purple), with associated parameters $(a_6^*,\sigma_6^*) = (1.45,0.50)$ for ${\tilde{\Lambda}}_6[\omega]$, $\omega_4^*$ (red), with $(a_4^*,\sigma_4^*) = (1.43,0.42)$ for ${\tilde{\Lambda}}_4[\omega]$, and $\omega_3^*$ (blue), with $(a_3^*,\sigma_3^*) = (1.38,0.10)$ for ${\tilde{\Lambda}}_3[\omega]$. In the legend, we present the values of the next-to-largest eigenvalue mentioned in the main text.
  }
  \label{fig:instabilities-fixed}
\end{figure}

We further support the above picture by applying the same methodology to the other two bi-parametric families of jump distributions: (i) the algebraic $(a,\alpha)$ one, Eq.~\eqref{eq:alpha-a-distribution}, and (ii) the box-like $(a,b)$ one, Eq.~\eqref{eq:a-b-distribution}. We recall that the algebraic and the box-like families tend to a double Dirac-delta peak in the limit $\alpha\to\infty$ and $b\to 0^+$, respectively. This implies that we have to impose cutoffs in $\alpha$ and $b$, respectively, to avoid convergence problems: specifically, we have employed the values $\alpha_{\cutoff} = 4$ and $b_{\cutoff}=0.2$. In Fig \ref{fig:optimal-ansatz}, we show again that the collapse instability arises when considering the diffusion-rejection estimator ${\tilde{\Lambda}}_3[\omega]$, and that this instability is fixed by incorporating the CDW basis, i.e. by considering, at least, ${\tilde{\Lambda}}_4[\omega]$~\footnote{Of course, more complex functionals like ${\tilde{\Lambda}}_6[\omega]$, which also include the CDW basis, also fix the instability and improve the estimate for the next-to-largest eigenvalue.}. When using the diffusion-rejection estimator ${\tilde{\Lambda}}_3[\omega]$, the optimal parameters in the distributions $\omega_3^*$ reach the corresponding cutoffs, thus showing the collapse instability. Going to the estimators ${\tilde{\Lambda}}_4[\omega]$ or ${\tilde{\Lambda}}_6[\omega]$ solves this issue again. In fact, the global picture is completely analogous to the one found for the family of Gaussian distributions: (i) in all cases, ${\tilde{\Lambda}}_3^*$, ${\tilde{\Lambda}}_4^*$ and ${\tilde{\Lambda}}_6^*$ are quite close---of course, verifying always ${\tilde{\Lambda}}_3^*<{\tilde{\Lambda}}_4^*<{\tilde{\Lambda}}_6^*$, (ii) as compared with ${\tilde{\Lambda}}_6^*$, ${\tilde{\Lambda}}_6[\omega_3^*]$ has a large error that is radically diminished by ${\tilde{\Lambda}}_6[\omega_4^*]$.
\begin{figure}
  \centering
  \includegraphics[width=3.1in]{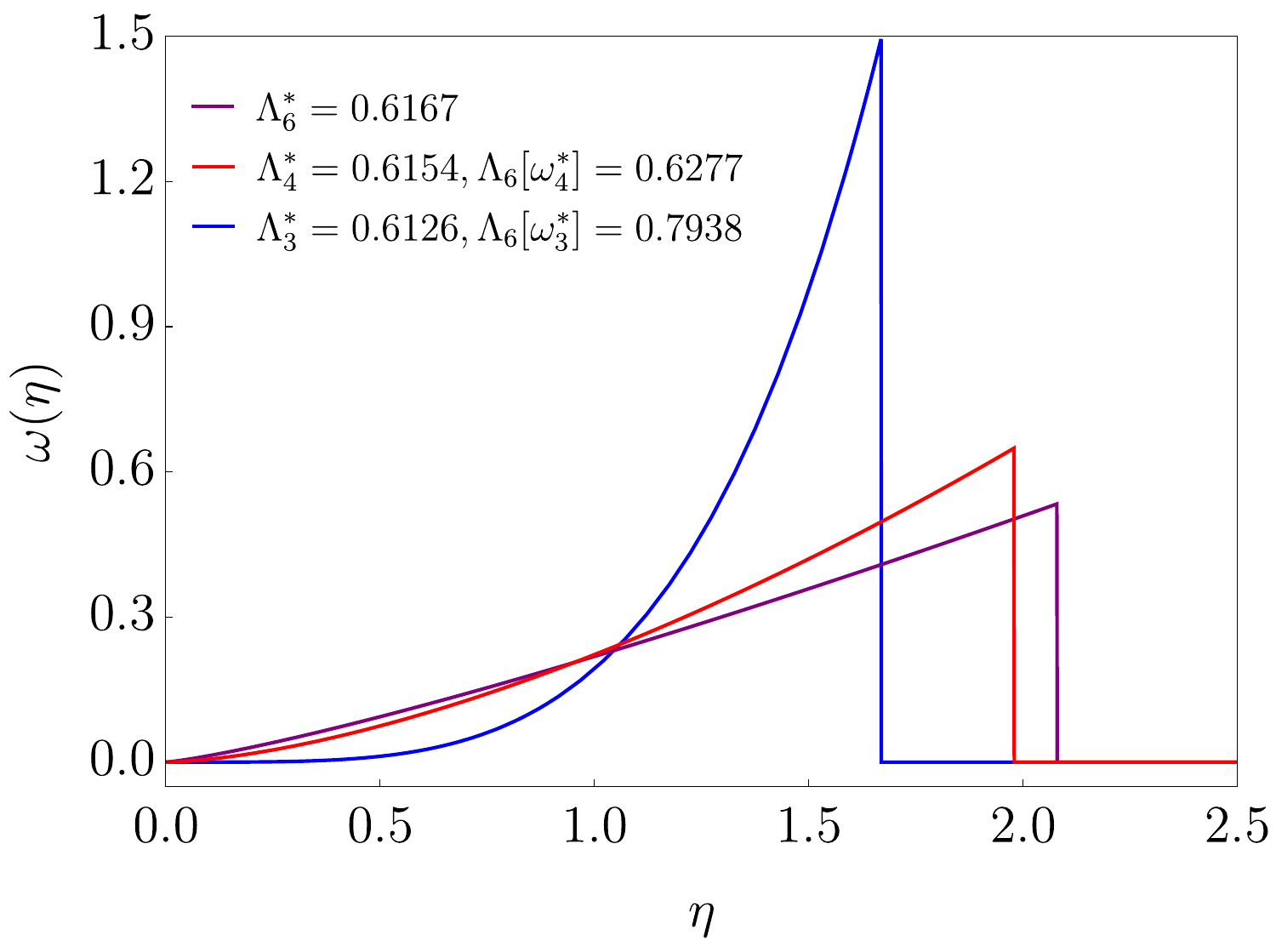}
  \includegraphics[width=3.1in]{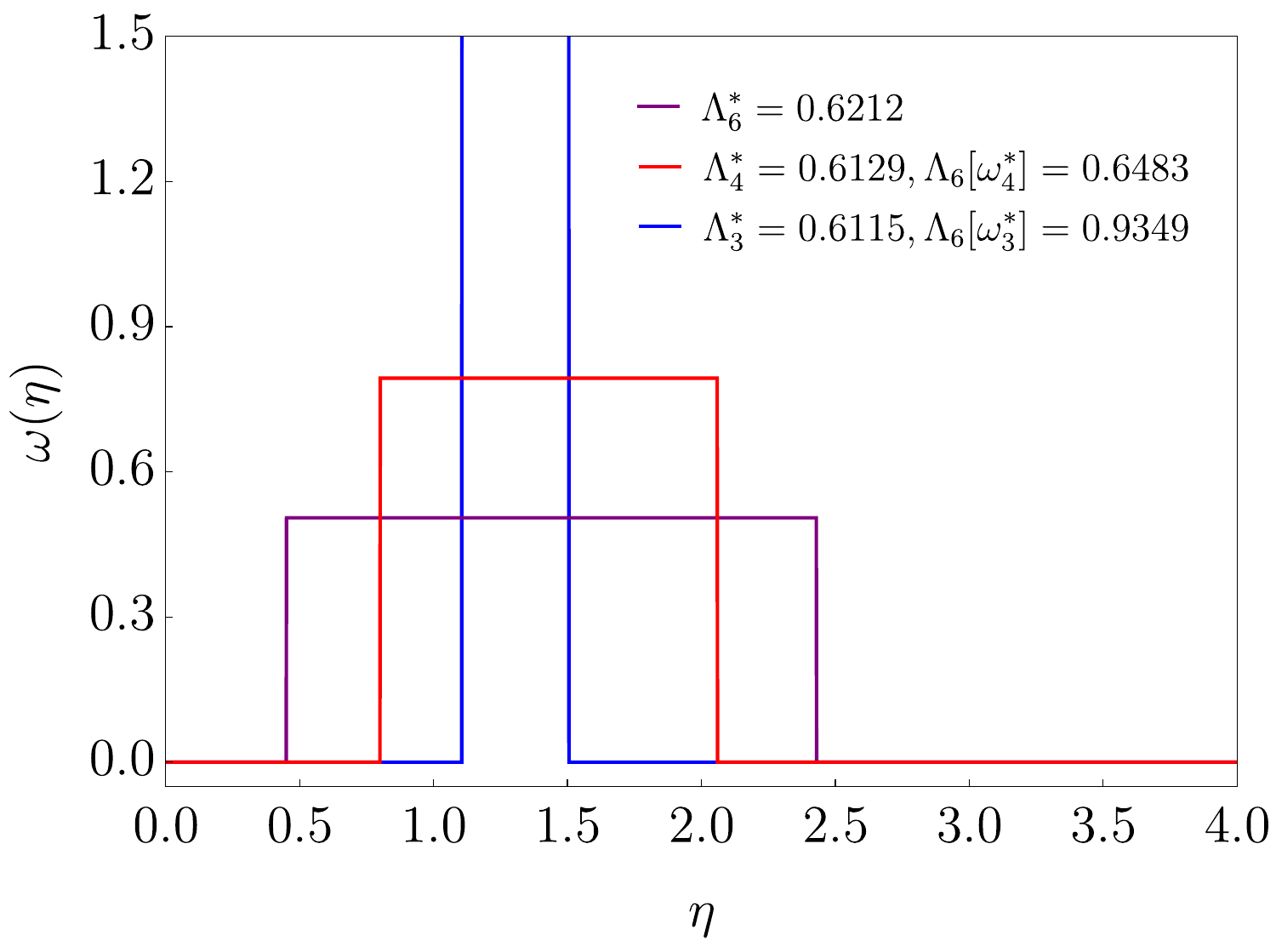}
  \caption{Optimal jump distributions for the algebraic and box-like family of distributions. Specifically, the algebraic \eqref{eq:alpha-a-distribution} and box-like \eqref{eq:a-b-distribution} cases are considered in the left and right panels, respectively. Similarly to Fig.~\ref{fig:instabilities-fixed}, the optimal jump distributions depicted correspond to the functionals ${\tilde{\Lambda}}_3[\omega]$, ${\tilde{\Lambda}}_4[\omega]$, and ${\tilde{\Lambda}}_6[\omega]$. Specifically, we show: $\omega_{6}^*$(purple), for which $(a_6^*,\alpha_6^*) = (2.08,1.22)$ and $(a_6^*,b_6^*) = (1.44, 0.99)$,  $\omega_{4}^*$ (red), for which $(a_4^*,\alpha_4^*) = (1.98,1.57)$ and $(a_4^*,b_4^*) = (1.43, 0.63)$ and (blue) $\omega_{3}^*$, for which $(a_3^*,\alpha_3^*) = (1.67,4.00)$ and $(a_3^*,b_3^*) = (1.30, 0.20)$. Again, the values of the {next-to-largest eigenvalue} mentioned in the main text are given in the legend.
  }
  \label{fig:optimal-ansatz}
\end{figure}

\section{On the shape of the optimal jump distribution}\label{sec:further-opt}

In the previous sections, we have carried out the minimisation of the next-to-largest eigenvalue over several bi-parametric families of jump distributions, all of which tend to a double Dirac-delta peak in a certain limit. Here, we carry out this minimisation over more complex jump distributions. In this way, we try to shed some light on the global optimisation problem, i.e. given a confining potential $U(x)$, look for the jump distribution $\omega^*(\eta)$ that maximises the convergence rate of the dynamics. {We highlight once again that the following analysis is carried out for the case of harmonic confinement $U(x) = \chi x^2/2$. In general, the shape of the optimal jump distribution depends on the form of the confining potential $U(x)$, and the harmonic potential just constitutes the paradigmatic example of a convex potential with a unique global minimum.}

\subsection{More complex jump distributions}\label{sec:more-complex}

In order to see if we can further improve our estimation of the global minimum value of the {next-to-largest eigenvalue},
we consider a more complex, discrete-point, family of the form:
\begin{equation}\label{eq:deltas}
    \omega(\eta) = \sum_{j=1}^{N_{\text{P}}}a_j \,\delta (|\eta| - \eta_j),
\end{equation}
where both the weights $a_j$ and the positions $\eta_j$ of each of the $N_{\text{P}}$ delta peaks are adjustable parameters, over which we carry out the optimisation procedure. Figure~\ref{fig:deltas} shows the optimal jump distribution for the case $N_{\text{P}}=20$, employing the ${\tilde{\Lambda}}_6[\omega]$ functional for the next-to-largest eigenvalue. With this approach, we get ${\tilde{\Lambda}}_6^*=0.6146$, which is slightly smaller than the previously found values for the two-parametric families in Sec.~\ref{sec:fix-collapse-ins}. But, more interestingly, although most of the optimal jump distribution is concentrated in a definite interval, $0.5\lesssim \eta \lesssim 2.0$, there are a couple of delta peaks for larger jumps, around $\eta=3$. This behaviour hints at the actual optimal jump distribution having a more complicated structure, with at least four peaks---recall that $\omega(\eta)$ is an even function of $\eta$, so the two peaks found for $\eta>0$ are replicated for $\eta<0$.
\begin{figure}
  \centering
  \includegraphics[width=3.3in]{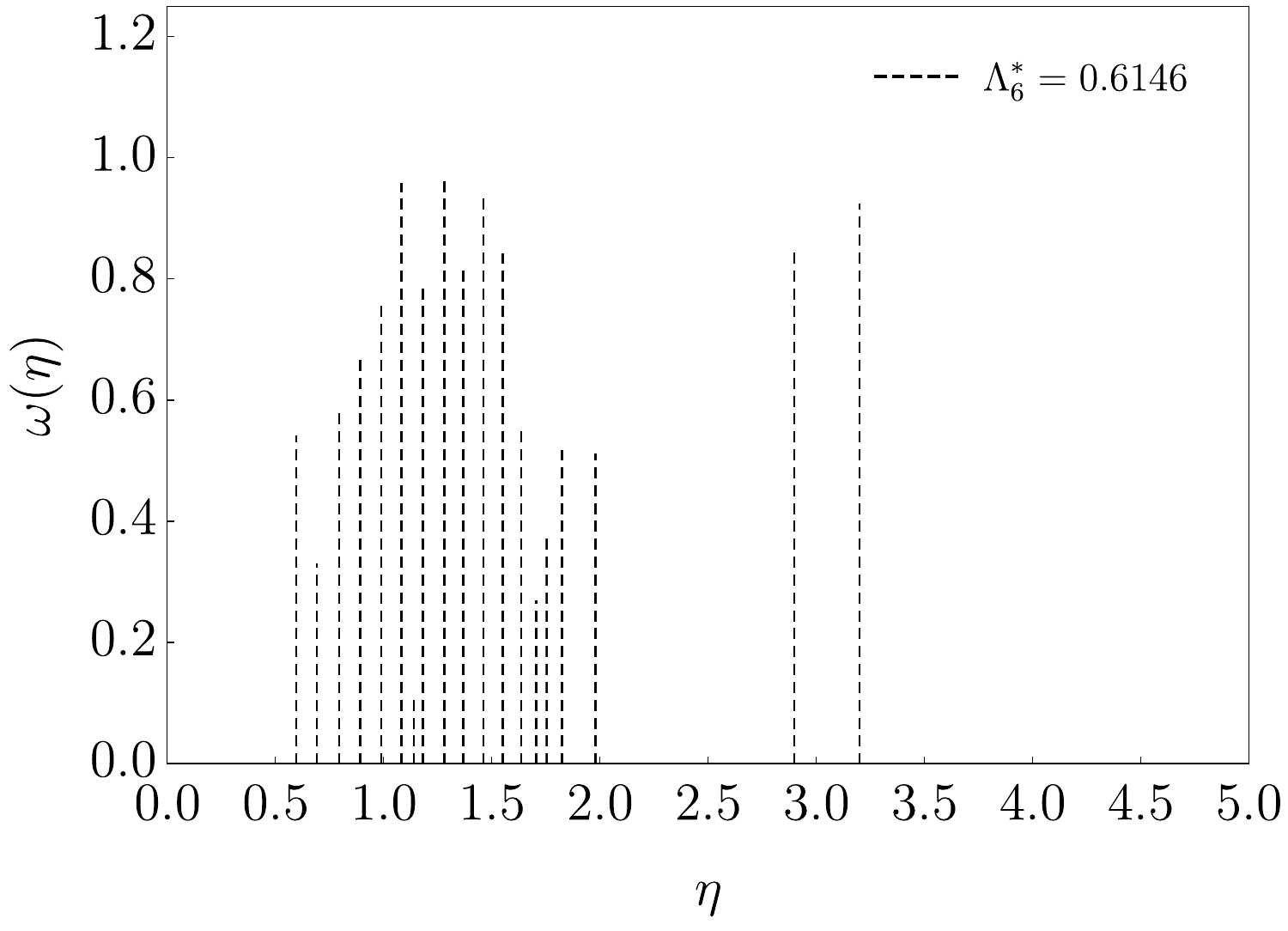}
  \caption{ Optimal jump distribution for the discrete-point family of Eq.~\eqref{eq:deltas}. Specifically, we have taken $N_{\text{P}} = 20$. The vertical dashed lines give the value of the coefficients $a_j$ of each of the delta peaks at discrete positions $\eta_j$. The optimal jump distribution has been obtained using the ${\tilde{\Lambda}}_6$ functional, and the corresponding estimate ${\tilde{\Lambda}}_6^*$ for the {next-to-largest eigenvalue} is shown in the legend.}
  \label{fig:deltas}
\end{figure}

To check the above insights, we have employed the following two-Gaussian family of jump distributions with a cutoff,
\begin{equation}\label{eq:two-gaussians-cutoff}
    \omega (\eta) = C \sum_{n=1}^2  \alpha_n \ \text{exp}\left(-\frac{(|\eta|-a_n)^2}{2\sigma_n^2}\right)\theta(\eta_{\max}-|\eta|), \quad a_1<a_2,
\end{equation}
where $C$ is a normalisation constant. Note that this family has seven parameters: $(a_1,{\alpha_1},\sigma_1,a_2,{\alpha_2},\sigma_2,\eta_{\max})$. The constants $\alpha_i>0$ give the relative weights of the four peaks at $\eta=\pm a_1$ and $\eta=\pm a_2$. In addition, this family includes as a particular case the bi-parametric Gaussian family in Eq.~\eqref{eq:parametric-gaussian}, in the limit $\eta_{\max}\to\infty$ and (e.g.) $\alpha_1\to 0^+$.
\begin{figure}
  \centering
  \includegraphics[width=3.3in]{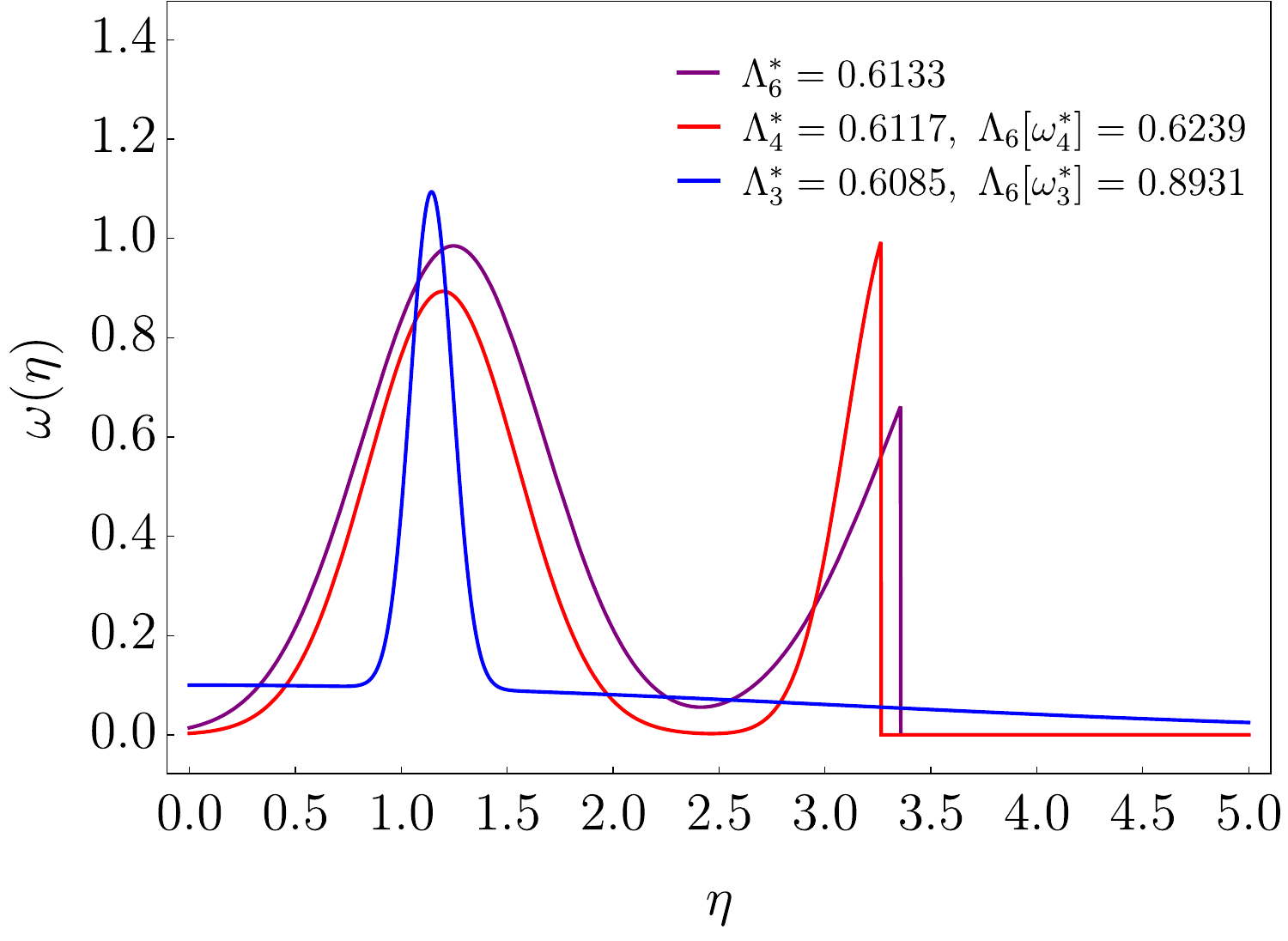}
  \caption{Optimal jump distribution for the 7-parameter family in Eq.~\eqref{eq:two-gaussians-cutoff}. Similarly to the previous figures, the optimal jump distributions have been obtained using three different functionals:  ${\tilde{\Lambda}}_6^*$ (purple),  ${\tilde{\Lambda}}_4^*$ (red), and  ${\tilde{\Lambda}}_3^*$ (blue). In the legend, we present the values of the {next-to-largest eigenvalue} mentioned in the text. The parameters characterising each jump distribution are shown in Table~\ref{table:gaussians}.}
  \label{fig:cutoff}
\end{figure}

Figure~\ref{fig:cutoff} shows the optimal jump distributions for the 7-parameter family~\eqref{eq:two-gaussians-cutoff}, once more employing the functionals ${\tilde{\Lambda}}_M[\omega]$ with $M=3$, $4$, and $6$. Again, the CDW fixes the collapse instability shown by the estimate $\omega_3^*$ of the diffusion-rejection estimator---note that $\sigma_1$ again has reached the cutoff value $\sigma_{\cutoff}=0.1$. Interestingly, this 7-parameter family slightly improves the optimal next-to-largest eigenvalue: here we have ${\tilde{\Lambda}}_6^* = 0.6133$, which is smaller than the values for the previous bi-parametric families, displayed in Figs. \ref{fig:instabilities-fixed}-\ref{fig:optimal-ansatz}. Apart from the small decrease in the optimal next-to-largest eigenvalue, the optimal jump distributions $\omega_4^*$ and $\omega_6^*$ have a four-peak structure, analogous to that found in Fig.~\ref{fig:deltas} for the discrete-point family---recall that we only plot $\omega(\eta)$ for $\eta>0$, due to its even parity.

The above analysis points out that the global optimal jump distribution may present at least four peaks, corresponding to jumps centred around $\pm a_1$ and $\pm a_2$, instead of the two peaks at $\pm a$ we have considered in the previous sections.
\begin{table}
\centering
 \begin{tabular}{|| P{1.0 cm} | P{1. cm} | P{1. cm}  | P{1. cm} | P{1. cm} | P{1. cm}| P{1. cm}| P{1. cm}|} 
 \hline
  & $\eta_{\max}$ & $\alpha_1$ & $a_1$ & $\sigma_1$ & $\alpha_2$ & $a_2$  & $\sigma_2$ \\ [1ex]
 \hline
 $\omega_6^*$ & $3.358$ & $0.985$ & $1.247$ & $0.428$ & $0.986$ & $3.845$  & $0.546$ \\ [1ex]
 \hline
 $\omega_4^*$ & $3.265$ & $0.893$ & $1.198$ & $0.354$ & $1.050$ & $3.343$  & $0.235$ \\ [1ex]
 \hline
 $\omega_3^*$ & $5.764$ & $1.000$ & $1.144$ & $0.100$ & $0.100$ & $0.074$  & $2.950$ \\ [1ex]
 \hline
\end{tabular}
\caption{Parameters characterising the optimal jump distributions for the 7-parameter family in Eq.~\eqref{eq:two-gaussians-cutoff}. As in previous cases, optimisation is carried out for the functionals ${\tilde{\Lambda}}_M[\omega]$, with $M=3$, $4$ and $6$.}
\label{table:gaussians}
\end{table}

\subsection{Sequence of jumps}\label{sec:sequence}

The fact that four-peak jump distributions improve the convergence rate as compared with two-peak jump distributions, uncovered in Sec.~\ref{sec:more-complex}, also provides insights into the design of other strategies for further optimising the convergence rate. As an example of such a possible strategy, we consider the deterministic alternation between two jump distributions $\omega_1(\eta)$ and $\omega_2(\eta)$ with peaks at $\eta=\pm a_1$ and $\eta=\pm a_2$, respectively. 

The corresponding kernel $\hat{\mathcal{K}}(\omega_1,\omega_2)$ for the master equation of this modified Metropolis algorithm is
\begin{equation}\label{kernel-3}
    \hat{\mathcal{K}}(\omega_1,\omega_2) = \hat{K}(\omega_1) \hat{K} (\omega_2)\hat{K}(\omega_2) \hat{K}(\omega_1),
\end{equation}
with $\hat{K}(\omega_i)$ being the kernel from our original Metropolis algorithm from Eq.~\eqref{kernel-2}, with a fixed choice $\omega_i(\eta)$ for the jump distribution. We are employing parenthesis instead of the brackets that appeared in Eq.~\eqref{kernel-2}, in order to highlight that $\omega$ is no longer a variational parameter. This new kernel describes a modified Metropolis algorithm in which two jumps from $\omega_1$ are attempted, then two from $\omega_2$ are attempted, and the process repeats. The sequence have been chosen so that the kernel form in Eq.~\eqref{kernel-3} remains symmetric. 

In the following, we numerically compare the efficiency of this strategy against {the} original Metropolis algorithm {studied in this work, but with the average} jump distribution $\omega = (\omega_1 + \omega_2)/2$. Now, as a consequence of the structure of the sequential kernel in Eq.~\eqref{kernel-3}, the corresponding {next-to-largest eigenvalue} for the modified Metropolis algorithm is $\Lambda_{\seq} = (\Lambda_1 \Lambda_2)^{1/2}$, with $\Lambda_i$ being the {next-to-largest eigenvalue} for the jump distribution $\omega_i$. As {shown in Fig.~\ref{fig:sequence},} the sequence of jumps improves the convergence time of the original algorithm, {by lowering the next-to-largest eigenvalue from $\Lambda_6^* = 0.6133$ to $\Lambda_{\seq} = 0.5840$. Since the convergence rate is given by $\gamma=|\ln\Lambda|$, as discussed in Sec.~\ref{sec:convergence-rate}, this entails that the actual speed-up obtained with a sequence of jumps is about $10$ per cent.} We remark that this improvement is just for a reduced number of degrees of freedom: in this case, one standard deviation and two mean values for the Gaussian ansatz. Increasing the dimension of the parameter space, or even combining different families of jump distributions---such as the ones employed before, given by Eqs.~\eqref{eq:parametric-gaussian}, \eqref{eq:alpha-a-distribution} and \eqref{eq:a-b-distribution}---should further improve the {optimisation} of the convergence rate. 
\begin{figure}
  \centering
  \includegraphics[width=3.3in]{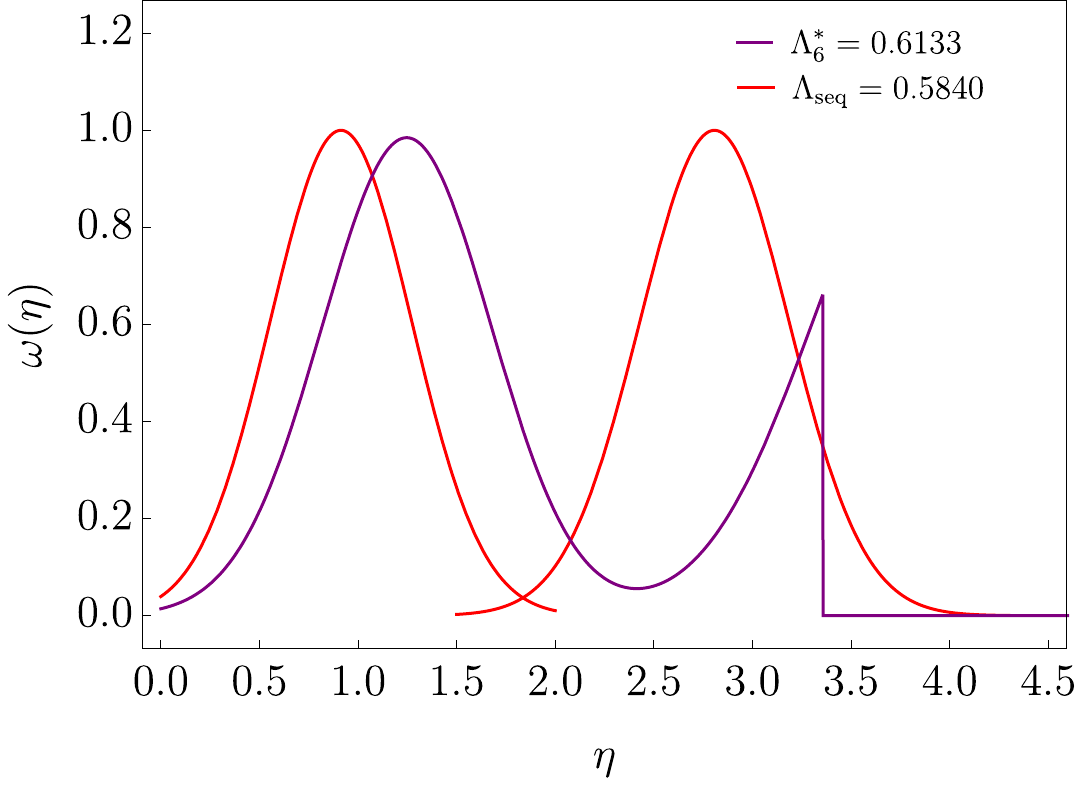}
  \caption{Comparison of optimal jump distributions. Specifically, we show the optimal jump distributions for (i) the 7-parametric jump distribution in Eq.~\eqref{eq:two-gaussians-cutoff}, already shown in Fig.~\ref{fig:cutoff} (purple), and (red) the sequence of jumps described in the text, alternating two Gaussian distributions with the same standard deviation but different mean value. The corresponding optimal {next-to-largest eigenvalues} mentioned in the text are also plotted in the legend.}
  \label{fig:sequence}
\end{figure}

We may understand the better performance of the sequence-of-jumps strategy over the original Metropolis scheme with the following intuitive argument. Let us take the characteristic lengths $a_1 < a_2$, with a sufficient gap between the two. Then, for $\omega_1(\eta)$, the first two steps of the algorithm will resemble that of free diffusion, which allows exploration of small regions of the phase space. In order to move further, we take the second two steps, which allow one to jump to further points in space, and thus the rejection rate increases accordingly. Thus, each distribution balances the problems presented for the other. 

The previous argument has remarkable implications in terms of choosing between stochastic and deterministic jump strategies. On the one hand, the former would correspond to our original Metropolis algorithm: if we consider the four-peak jump distribution $\omega (\eta)$ as the average of two independent distributions $\omega_1(\eta)$ and $\omega_2(\eta)$ with different characteristic lengths, $\omega=(\omega_1+\omega_2)/2$, we may understand the algorithm as choosing randomly between these two distributions before attempting any jumps. On the other hand, the deterministic approach would correspond to the sequence of jumps, since we clearly set the order in which we draw the attempted jumps from $\omega_1$ and $\omega_2$.

\section{Conclusions}\label{sec: conclusions}

In this paper, we have analysed the relaxation dynamics of the Metropolis algorithm for a particle confined in a one-dimensional potential, with even jump distributions that ensure that the dynamics verify detailed balance. The central result of this paper is the functional optimisation of the convergence rate of the Metropolis algorithm in one dimension, which is obtained as a balance of three mechanisms: diffusion, rejection, and the novel CDW regime that we unveil and characterise in this paper. In the CDW regime, the leading eigenfunction $\phi_1(x)$ corresponding to the {next-to-largest eigenvalue $\Lambda\equiv\lambda_1$---we recall that the largest eigenvalue is the equilibrium one} $\lambda_0=1$---shows an oscillatory behaviour, characterised by a wave number $k$. For two-parametric jump distributions, the balance among diffusion, rejection, and CDW occurs at the tricritical point of coexistence---at which the minimum value of the {$\Lambda$, i.e. the maximum value of the convergence rate $\gamma=|\ln\Lambda|$} is attained.

{The first of our core results constitutes completing the physical picture of the dominant mechanisms driving the relaxation dynamics of the Metropolis algorithm. The balance between diffusion and rejection makes it possible to obtain the optimal convergence rate for several choices of confining potentials and jump distributions~\cite{chepelianskii_metropolis_2023}, but it does not explain the eigenvalue crossing taking place for sufficiently narrow jump distributions. Our numerical and analytical approaches show that such crossing points out towards a first-order phase transition between the diffusion and the---newly characterised---CDW regime, which involves a change of symmetry in the dominant eigenmodes.}

This interplay between {the aforementioned} physical factors has been illustrated by the phase diagram for the {next-to-largest eigenvalue $\Lambda$}. Specifically, this phase diagram has been built for the algebraic $(a,\alpha)$ family of jump distributions of Eq.~\eqref{eq:alpha-a-distribution}. The different phases are characterised by two order parameters: fidelity $\calF$, which is different from zero when the new CDW regime dominates, and IPR (inverse participation ratio), which is different from zero when the rejection regime dominates. First, for smooth jump distributions without a double-peak structure, we have the localisation transition between the diffusion and rejection regions as the length of the typical jumps is increased---which was characterised in Ref.~\cite{chepelianskii_metropolis_2023}. Then, we have unveiled a new transition between the diffusion and CDW regimes in the parameter range for which the jump distribution approaches a double peak structure, also as the length of the typical jumps is increased. The diffusion-CDW transition is first-order, so that (i) $\calF$ and the optimal wave number abruptly change, and (ii) the parity of the leading eigenfunction of the master equation changes---from odd to even. Finally, it is only when the length of the typical jumps is further increased that we find a transition between the CDW and rejection regimes. At this transition, both order parameters vary smoothly, so the CDW-rejection transition is at least second order. 

{As our second major result, we have searched for the functional optimal jump distribution (meaning that no a priori assumption is made on the functional shape), so that the relaxation dynamics is the shortest possible. We have found the corresponding jump distribution to present four peaks,
rather than two as studied above.
Besides, 
}we have shown numerical evidence on the improvement of the convergence rate by combining different jump distributions. The sequence of four jumps exposed in this work is just an example of how the interplay between more than one characteristic length may improve the relaxation to equilibrium. Interestingly, a similar improvement of other figures of merit when considering more than one characteristic length has been observed in other physical systems---e.g. for optimal search strategies and resetting processes~\cite{benichou_intermittent_2011,evans_stochastic_2020,garcia-valladares_optimal_2023}. 

{The variational method employed to estimate the optimal convergence rate is quite general and could, in principle, be extended to study optimisation problems in higher dimensions or in systems with multiple minima. However, it is important to consider the computational challenges associated with such extensions. Specifically, (i) the number of integrals in the master equation increases with each additional dimension, significantly raising computational complexity; (ii) the convergence of the method may be compromised, as a higher number of eigenmodes from the truncated basis would likely be required to achieve accurate estimations in higher-dimensional systems; and (iii) for potentials with multiple minima, a larger and more diverse basis set might be necessary to adequately capture the system's behaviour. Despite these challenges, such generalisations represent a promising avenue for future work.}

In this paper, we have completely characterised the CDW regime in { the one-dimensional case. Therein, an isotropic jump distribution with a fixed jump size, i.e. the Dirac-delta jump distribution in Eq. \eqref{eq:omega0-delta-peak}, does not allow the system to explore the whole configuration space. It is this ergodicity---or irreducibility---breaking that gives rise to the CDW behaviour.} Thus, the natural question is whether or not the CDW regime would play a role in the global optimisation of the Metropolis algorithm in higher dimensions. {Intuitively, it seems plausible that the role of the CDW regime in the optimisation problem may be less relevant in higher-dimensional systems.  This stems from isotropic jump distributions with a radial Dirac-delta component not inherently breaking ergodicity: the whole configuration space can be explored with a sequence of fixed-size jumps---already in two dimensions. Consequently, the influence of the CDW regime on the convergence properties of the Metropolis algorithm might diminish with dimensionality. Nonetheless, a detailed investigation is required to check the accuracy of the intuitive picture above and to determine whether or not higher-dimensional analogues of the CDW regime could emerge under specific conditions.}

Additionally, it is relevant to investigate the optimisation of the convergence rate for other dynamics, different from the Metropolis algorithm---{such as the} Glauber {algorithm for simulating the Ising Model}~\cite{glauber_time-dependent_1963}{, or the Simulated Annealing techniques~\cite{kirkpatrick_optimization_1983,aarts_simulated_1989,van_laarhoven_simulated_1987}, which incorporate the Metropolis algorithm with a time-dependent cooling program for the temperature $T$ in order to attain the global minimum of a function}. We would like to stress that our analysis has focused on the leading eigenvalue $\lambda_1$, which as a rule controls the convergence rate to equilibrium. However, there are certain specific situations in which the convergence rate to equilibrium is controlled by the following eigenvalue $\lambda_2<\lambda_1$, because the coefficient of the first mode vanishes, as in the strong version of the Mpemba effect~\cite{lu_nonequilibrium_2017,klich_mpemba_2019,kumar_exponentially_2020,busiello_inducing_2021,kumar_anomalous_2022,zhang_theoretical_2022,biswas_mpemba_2023,patron_non-equilibrium_2023} or when specific strategies are employed to tune the dynamical evolution of the system~\cite{gal_precooling_2020,chittari_geometric_2023}. Therefore, it is also worth investigating whether the general picture found here for $\lambda_1$, i.e., the competition between diffusion, rejection, and CDW regimes, extends to $\lambda_2$ or even to the full eigenvalue spectrum of the master equation. Certainly, all these questions open the door to future research.

\section{Data Availability Statement}

The codes employed for generating the data that support the findings of this study, together with the Mathematica notebooks employed for producing the figures presented in the paper, are openly available in the~\href{https://github.com/fine-group-us/optimal_convergence_rate_metropolis}{GitHub page} of University of Sevilla's FINE research group.

\section{Acknowledgements}

We thank Satya Majumdar for fruitful discussions. A. Patrón and A. Prados acknowledge financial support from Grant PID2021-122588NB-I00 funded by MCIN/AEI/ 10.13039/501100011033/ and by ``ERDF A way of making Europe''. A. Patrón, A. Prados, and E. Trizac acknowledge financial support from Grant ProyExcel\_00796 funded by Junta de Andalucía's PAIDI 2020 programme. A. Patr\'on also acknowledges support from the FPU programme through Grant FPU2019-4110, and  additional support from the FPU programme through Grant EST22/00346, which funded his research stay at Univ.~Paris-Saclay during autumn 2022. A.~Prados also acknowledges the hospitality of LPTMS, which funded his stay at Univ.~Paris-Saclay in June 2022. A. Chepelianskii acknowledges financial support from Grant ANR-20-CE92-0041 (MARS).

\appendix

\section{Analytical derivation of the CDW eigenfunctions}\label{app:analytical-derivation}

We start by analysing the second term on the {lhs} of Eq.~\eqref{eq:eigenproblem-2-b-explicit}, which involves the rejection probability defined in Eq.~\eqref{eq:rejection-prob}. By defining $\ell(x)$ and $h(x)$ as the $x'$-values at which $U(x')=U(x)$, the Heaviside function restricts the integral over $x'$ to a certain interval $(-\infty,\ell(x))\cup(h(x),+\infty)$. Depending on the sign of $U'(x)$, either $\ell(x)$ or $h(x)$ equals $x$---in the following we focus on the latter case, $U'(x)>0$ ($x>0$), where $h(x)=x$. With these definitions,
\begin{equation}
    R(x;a,\sigma)=
    \left(\int_{-\infty}^{\ell(x)}+\int_{x}^{+\infty} \right) dx' \omega(x'-x;a,\sigma) \left[1 - e^{-\beta {\Delta U(x',x)}}\right].
\end{equation}
The first integral is subdominant, since $\omega$ is basically zero over the considered interval, as long as $\ell(x)-x<-a$, i.e. $x-a>\ell(x)$. On the other hand, in the second integral we have that $x'-x>0$ and therefore only one of the $\delta_{\sigma}$'s contribute:
\begin{equation}\label{eq:rej-05}
    R(x;a,\sigma)\approx \frac{1}{2} \int_{x}^{+\infty} dx' \delta_{\sigma}(x'-x-a) \left[1 - e^{-\beta {\Delta U(x',x)}}\right]\approx \frac{1}{2} \left[1-e^{-\beta {\Delta U(x+a,x)}}\right]\approx \frac{1}{2}, 
\end{equation}
assuming that 
\begin{equation}\label{eq:exp-factor-small}
   e^{-\beta {\Delta U(x+a,x)}}\ll 1, 
\end{equation}  
{which is analogous to Eq.~\eqref{eq:ansatz-1} from the main text.} This condition implies that $a$ is large enough, see Appendix~\ref{app:robustness-cdw} for more details. {In fact, the relationship between Eqs.~\eqref{eq:rej-05} and ~\eqref{eq:exp-factor-small} follows from {the} intuitive physical arguments {presented in Sec.~\ref{subsec:main-assumptions}}: If $a$ is sufficiently large and $x$ not too close to zero---i.e. the global minimum, such that the potential energy increases significantly, then most of the jumps to the right will be rejected, while those to the left are basically accepted. Thus, the overall rejection probability is close to $1/2$.} Now we move on to analysing the first term on the {lhs} of Eq.~\eqref{eq:eigenproblem-2-b-explicit}, splitting the integral into two slices: one in which $U(x) > U(x')$, i.e., the interval $(\ell(x),x)$ and another in which $U(x)< U(x')$, i.e. $(-\infty,\ell(x))\cup(x,+\infty)$. For the first slice, $U(x) > U(x')$, we have 
\begin{align}
    \int_{-\infty}^{+\infty}dx' & \theta({\Delta U(x,x')})\,\omega(x-x';a,\sigma)\,e^{-\beta |{\Delta U(x,x')}|/2}\rho_{\nu}(x')e^{-\beta U(x')/2} \nonumber \\
    &=\int_{\ell(x)}^{x}dx'\omega(x-x';a,\sigma) e^{-\beta (U(x)-\cancel{U(x')})/2}\rho_{\nu}(x')e^{-\beta \cancel{U(x')}/2} \approx \frac{1}{2}e^{-\beta U(x)/2}  \int_{\ell(x)}^{x}dx'\delta_{\sigma}(x-x'-a) \rho_{\nu}(x') \nonumber \\
    &\approx \frac{1}{2} e^{-\beta U(x)/2} \int_{-\infty}^{+\infty}dx'\delta_{\sigma}(x-x'-a) \rho_{\nu}(x')=\frac{e^{-\beta U(x)/2}}{2}  \int_{-\infty}^{+\infty}dx'\delta_{\sigma}(x-x') \rho_{\nu}(x'-a) \nonumber \\
    &\approx \frac{e^{-\beta U(x)/2}}{2} \int_{-\infty}^{+\infty}dx'\delta_{\sigma}(x-x') \left[\rho_{\nu}(x-a)+\frac{1}{2}\rho_{\nu}''(x-a) (x-x')^2 \right] \nonumber \\ 
    &=\frac{1}{2} e^{-\beta U(x)/2}\left[\rho_{\nu}(x-a)+\frac{\sigma^2}{2}\rho_{\nu}''(x-a) \right],
    \label{good-terms}
\end{align}
where we have taken into account again that (i) only one of the $\delta_{\sigma}$'s contribute, and (ii) the interval of integration may be extended from $(\ell(x),x)$ to $(-\infty,+\infty)$, once the peak of the  $\delta_{\sigma}$ is included in the considered interval---again, as long as $x-a >\ell(x)$. For the second slice, $U(x') > U(x)$, we have 
\begin{align}
    \int_{-\infty}^{+\infty}dx' & \theta({\Delta U(x',x)})\,\omega(x-x';a,\sigma)\,e^{-\beta |{\Delta U(x,x')}|/2}\rho_{\nu}(x')e^{-\beta U(x')/2} \nonumber \\
    &=\left(\int_{-\infty}^{\ell(x)}+\int_{x}^{+\infty} \right) dx'\omega(x-x';a,\sigma) e^{-\beta {\Delta U(x',x)}/2}\rho_{\nu}(x')e^{-\beta U(x')/2} \nonumber \\
    &\approx \frac{e^{\beta U(x)/2}}{2}  \int_{x}^{+\infty}dx'\delta_{\sigma}(x+a-x') \rho_{\nu}(x') e^{-\beta U(x')} \approx \frac{e^{\beta U(x)/2}}{2}  \int_{-\infty}^{+\infty}dx'\delta_{\sigma}(x+a-x') \rho_{\nu}(x') e^{-\beta U(x')} \nonumber \\
    &\approx e^{-\beta {\Delta U(x+a,x)}} \frac{1}{2} e^{-\beta U(x)/2}\rho_{\nu}(x+a) +O(\sigma^2).
    \label{bad-terms}
\end{align}
Note that Eq.~\eqref{bad-terms} is subdominant against Eq.~\eqref{good-terms}, due to the presence of the exponential factor \eqref{eq:exp-factor-small}, so it is neglected in the following. The above analysis implies that the eigenvalue equation \eqref{eq:eigenproblem-2-b-explicit} can be approximated by
\begin{equation}\label{eq:CDW-approx-without-simplify}
\frac{1}{2} \cancel{e^{-\beta U(x)/2}}\left[\rho_{\nu}(x-a)+\frac{\sigma^2}{2}\rho_{\nu}''(x-a) \right]+\frac{1}{2} \cancel{e^{-\beta U(x)/2}} \rho_{\nu}(x)=\lambda_{\nu} \cancel{e^{-\beta U(x)/2}} \rho_{\nu}(x), 
\end{equation}
where $U'(x)>0$, i.e. $x>0$, provided that $x-a>\ell(x)$. A similar analysis for $U'(x)<0$ leads to a completely analogous expression for $x<0$, with the change $a\to -a$, provided that $x+a<h(x)$---note that $\ell(x)=x$ for $x<0$.

Summarising our findings, in the limit $\sigma\ll a$ we have the approximate equations
\begin{subequations}\label{eq:CDW-x-pos-neg}
\begin{equation}\label{eq:CDW-x-pos-neg-a}
\frac{\sigma^2}{4}\rho_{\nu}''(x-a)+\frac{1}{2}\Big[\rho_{\nu}(x-a)+\rho_{\nu}(x)\Big]=\lambda_{\nu} \rho_{\nu}(x), \quad x>0, \; x-a>\ell(x);
\end{equation}
\begin{equation}\label{eq:CDW-x-pos-neg-b}
\frac{\sigma^2}{4}\rho_{\nu}''(x+a)+\frac{1}{2}\Big[\rho_{\nu}(x+a)+\rho_{\nu}(x)\Big]=\lambda_{\nu} \rho_{\nu}(x), \quad x<0, \; x+a<h(x).
\end{equation}
\end{subequations}
For even potentials, $\ell(x)=-x<0$ for $x>0$ ($h(x)=-x>0$ for $x<0$). Then, for even potentials, Eq.~\eqref{eq:CDW-x-pos-neg-a} (Eq.~\eqref{eq:CDW-x-pos-neg-b}) is limited to $x>a/2$ ($x<-a/2$). A more careful analysis would be needed in the region $-a/2<x<a/2$ for even potentials---and a certain region of small $x$ for potentials without symmetry. This could have been anticipated after the approximations made in Eq.~\eqref{good-terms}: whereas the starting expression clearly vanishes in the limit as $x\to 0$, the final approximate expression does not.

As a check of the consistency of Eqs.~\eqref{eq:CDW-x-pos-neg}, note that a uniform solution corresponds to the equilibrium eigenvalue $\lambda_0=1$. Moreover, {as mentioned in Sec.~\ref{sec:CDW} from the main text,} for infinitely sharp jump distributions, i.e. in the case $\sigma=0$, the Metropolis algorithm loses ergodicity and any function $\phi_{\nu}(x)$ of the form \eqref{eq:from-phi-to-rho} with $\rho_{\nu}(x)$ being periodic with period $a$ is expected to be a stationary solution of the master equation, i.e. an eigenfunction corresponding to the eigenvalue $\lambda_0=1$. Note that this is indeed predicted by Eqs.~\eqref{eq:CDW-x-pos-neg}, as setting $\lambda_{\nu}=1$ for $\sigma=0$ yields $\rho_{\nu}(x\pm a)=\rho_{\nu}(x)$. Interestingly, Eqs.~\eqref{eq:CDW-x-pos-neg} still admit spatially periodic solutions, with period $a$, for $\sigma\ne 0$. { Hence, by introducing the CDW ansatz \eqref{eq:cdw-ansatz}, both Eqs.~\eqref{eq:CDW-x-pos-neg} simplify to
\begin{equation}
    \label{diff-eq-2}
    \rho_{\nu}''(x) + \frac{4(1-\lambda_{\nu})}{\sigma^2}\rho_{\nu}(x) = 0, \quad \forall x,
\end{equation}
corresponding to Eq.~\eqref{diff-eq} from the main text.}

\section{Robustness of the CDW eigenfunctions}\label{app:robustness-cdw}

{We stress the main assumptions employed above to derive the novel CDW regime. First, we have assumed that the potential is convex, $U''(x)>0$, so that it has only one minimum. As already stated, this assumption has been introduced to simplify the presentation but it is not essential: the result applies to a generic confining potential in one dimension, which may have multiple minima---the proof is lengthy and this is the reason why we do not present it here. {The spirit of the proof for a general potential $U(x)$ lies in the fact that, beyond the lengths $l(x)$ and $h(x)$, additional $x'$-values satisfying $U(x') = U(x)$ may arise due to the presence of multiple minima. In such a scenario, the integrals over $x'$ in Eq.~\eqref{eq:eigenproblem-2-b-explicit} would be split into several intervals. Depending on the sign of $U'(x)$ and the position $x$, either the leftmost or rightmost intervals could be handled using the same reasoning as in Appendix~\ref{app:analytical-derivation}. By carefully selecting assumptions analogous to those in Eqs.~\eqref{eq:ansatz-1} and \eqref{eq:exp-factor-small}, the additional integral terms could be neglected in favour of the dominant ones, ultimately leading to Eq.~\eqref{diff-eq}. However, these additional assumptions would further restrict the range of $a$ and $x$ values that support the CDW regime. The latter analysis is highly dependent on the explicit form of $U(x)$ and the number of minima, which justifies our choosing of a convex potential with a unique global minimum as a paradigmatic example to highlight the main feature behind the emergence of the CDW regime.}

Second, we have assumed that, in the regions where $U'>0$, the potential increases sufficiently fast, as expressed by Eq.~\eqref{eq:exp-factor-small}. Although this assumption might seem rather restrictive, we have shown in Sec.~\ref{sec:numerical-check-CDW} for the paradigmatic example of harmonic confinement, that it is not so. As is often the case with singular perturbation schemes, the small parameters do not need to be very small for their theoretical predictions be accurate. }

First, it must be stressed that Eq.~\eqref{eq:CDW-x-pos-neg-a} is only valid for $x>a/2$ and Eq.~\eqref{eq:CDW-x-pos-neg-b} for $x<-a/2$, and we have extended the equation to all $x$. Then, we have that $a$ must be ``small" enough for the CDW to emerge. On the other hand, Eq.~\eqref{eq:exp-factor-small} entails that $a$ must be ``large" enough. Since $U''(x)>0$, for $x>0$ one has
\begin{equation}
    1\gg e^{-\beta{\Delta U(x+a,x)}}=e^{-\beta a U'(\tilde{x})}>e^{-\beta a U'(x+a) },
\end{equation}
where $\tilde{x}\in[x,x+a]$. The convexity condition implies that it is enough to ensure the above relation for $x=0$, i.e. it is enough to have that $e^{-\beta a U'(a)}\ll 1$. For example, in the case of the harmonic potential, this is equivalent to having $e^{-\beta\chi a^2}\ll 1$. 

Second, we may consider the higher-order terms in the width of the distribution that have been neglected in Eq.~\eqref{good-terms}. This would result in the inclusion of the $O(\sigma^4)$ contribution into the term in brackets in Eq.~\eqref{eq:CDW-approx-without-simplify}, i.e.
\begin{equation}\label{eq:CDW-approx-without-simplify-higher-order}
\frac{1}{2} \cancel{e^{-\beta U(x)/2}}\left[\rho_{\nu}(x-a)+\frac{\sigma^2}{2}\rho_{\nu}''(x-a)+\xi \frac{\sigma^4}{4!} \rho_{\nu}''''(x-a)\right]+\frac{1}{2} \cancel{e^{-\beta U(x)/2}} \rho_{\nu}(x)=\lambda_{\nu} \cancel{e^{-\beta U(x)/2}} \rho_{\nu}(x), 
\end{equation}
where $\xi$ is a order of unity parameter, which gives the ratio between the fourth central moment of $\delta_{\sigma}$ and its variance squared, i.e.
\begin{equation}
    \int_{-\infty}^{+\infty} dz\, z^4 \delta_{\sigma}(z)\equiv \xi \sigma^4.
\end{equation}
Therefore, Eqs.~\eqref{eq:CDW-x-pos-neg} would be generalised to
\begin{subequations}\label{eq:CDW-x-pos-neg-hi-order}
\begin{equation}\label{eq:CDW-x-pos-neg-a-hi-order}
\xi\frac{\sigma^4}{48}\rho_{\nu}''''(x-a)+\frac{\sigma^2}{4}\rho_{\nu}''(x-a)+\frac{1}{2}\Big[\rho_{\nu}(x-a)+\rho_{\nu}(x)\Big]=\lambda_{\nu} \rho_{\nu}(x), \quad x>a/2;
\end{equation}
\begin{equation}\label{eq:CDW-x-pos-neg-b-hi-order}
\xi\frac{\sigma^4}{48}\rho_{\nu}''''(x+a)+\frac{\sigma^2}{4}\rho_{\nu}''(x+a)+\frac{1}{2}\Big[\rho_{\nu}(x+a)+\rho_{\nu}(x)\Big]=\lambda_{\nu} \rho_{\nu}(x), \quad x<-a/2.
\end{equation}
\end{subequations}
Along the same line of reasoning followed before, it is shown that we have periodic solutions with period $a$. It is only the dispersion relation that changes, i.e. Eq.~\eqref{eigen-CDW} holds but now $\lambda_{\nu}=1-\sigma^2 k_{\nu}^2/4 +\xi \sigma^4 k_{\nu}^4/48$. See also {Appendix~\ref{app:ergo-break}} for a further check of the robustness of the emergence of periodic solutions.

\section{Ergodicity breaking and periodic solutions}\label{app:ergo-break}

In this section, we provide more details on the emergence of periodic solutions in the limit $\sigma\to 0^+$. In this limit, the jump distribution reduces to the double Dirac-delta peak in Eq.~\eqref{eq:omega0-delta-peak}.
To be consistent with the discussion on the robustness of the CDW modes from Sec.~\ref{sec:numerical-check-CDW}, we restrict ourselves to even potentials $U(x)$, although a similar analysis can be carried out for a non-symmetric potential.

For $\sigma=0$, the eigenvalue/eigenfunction equation for the modes of the master equation \eqref{eq:eigenproblem-2-b-explicit} reduces exactly to
\begin{align}
     e^{-\beta|{\Delta U(x,x-a)}|/2}\phi_{\nu}(x-a)&+e^{-\beta|{\Delta U(x,x+a)}|/2}\phi_{\nu}(x+a)+
    {2-p(x-a,x)-p(x+a,x)}=2\lambda_{\nu}\phi_{\nu}(x), \quad \forall x.
\end{align}

It is clearer to split the above equation into three intervals, (i) $x>a/2$, (ii) $-a/2<x<a/2$, and (iii) $x<-a/2$. In this way, $U(x)-U(x\pm a)$ has a definite sign on each interval, and we get
\begin{subequations}\label{eq:sigma0-for-phinu}
    \begin{align}
        e^{\beta U(x-a)/2}\phi_{\nu}(x-a)+e^{-\beta {\Delta U(x+a,x)}} &\left[e^{\beta U(x+a)/2}\phi_{\nu}(x+a)-e^{\beta U(x)/2}\phi_{\nu}(x)\right] \nonumber
        \\
        &=(2\lambda_{\nu}-1) e^{\beta U(x)/2}\phi_{\nu}(x), & x>a/2, \\
        e^{-\beta {\Delta U(x+a,x)}}\left[e^{\beta U(x+a)/2}\phi_{\nu}(x+a)\right.&\left. -e^{\beta U(x)/2}\phi_{\nu}(x)\right] \nonumber \\
        +\ 
        e^{-\beta {\Delta U(x-a,x)}}\left[e^{\beta U(x-a)/2}\phi_{\nu}(x-a)\right.&\left. -e^{\beta U(x)/2}\phi_{\nu}(x)\right]\nonumber
        \\&=2(\lambda_{\nu}-1)e^{\beta U(x)/2}\phi_{\nu}(x), &  -a/2 < x <a/2, \\
        e^{\beta U(x+a)/2}\phi_{\nu}(x+a)+e^{-\beta {\Delta U(x-a,x)}}&\left[e^{\beta U(x-a)/2}\phi_{\nu}(x-a)-e^{\beta U(x)/2}\phi_{\nu}(x)\right] \nonumber
        \\
        &=(2\lambda_{\nu}-1) e^{\beta U(x)/2}\phi_{\nu}(x), & x<-a/2.
    \end{align}
\end{subequations}
Indeed, the structure of this equation suggests the change of variables in Eq.~\eqref{eq:from-phi-to-rho}. Introducing it into Eq.~\eqref{eq:sigma0-for-phinu},
\begin{subequations}\label{eq:sigma0-for-rhonu}
    \begin{align}
         \rho_{\nu}(x-a)+e^{-\beta {\Delta U(x+a,x)}}\left[\rho_{\nu}(x+a)\right.\left.-\rho_{\nu}(x)\right]&=(2\lambda_{\nu}-1) \rho_{\nu}(x), & x>a/2, 
         \label{eq:sigma0-for-rhonu-a}
         \\
         e^{-\beta {\Delta U(x+a,x)}}\left[\rho_{\nu}(x+a)-\rho_{\nu}(x)\right]+
        e^{-\beta {\Delta U(x-a,x)}}[\rho_{\nu}(x-a)-&\rho_{\nu}(x)]\nonumber
        \\
        &=2(\lambda_{\nu}-1)\rho_{\nu}(x), &  -a/2 < x <a/2, \\
        \rho_{\nu}(x+a)+e^{-\beta {\Delta U(x-a,x)}}\left[\rho_{\nu}(x-a)\right.\left.-\rho_{\nu}(x)\right]&=(2\lambda_{\nu}-1) \rho_{\nu}(x), & x<-a/2.
        \label{eq:sigma0-for-rhonu-c}
    \end{align}
\end{subequations}

It is worth stressing that Eq.~\eqref{eq:sigma0-for-rhonu} for $\rho_{\nu}(x)$ is exact for $\sigma=0$. Interestingly, any periodic function $\rho_{\nu}(x)$ with period $a$, i.e. $\rho_{\nu}(x\pm a)=\rho_{\nu}(x)$ $\forall x$, satisfies this equation with eigenvalue $\lambda_{\nu}=1$. As already discussed in the main text, this is intimately related to the breaking of ergodicity of the one-dimensional Monte Carlo algorithm for $\sigma=0$: any point $x$ is only connected to $x_n=x\pm n a$, $n \in\mathbb N$, and therefore the probability density of this lattice becomes distributed following the Boltzmann weight, i.e. it is proportional to $\exp(-\beta U(x_n))$, but the lattices corresponding to different values of $x$ are disconnected and the system does not reach equilibrium.

The consideration of a small width $\sigma$ makes it possible to introduce an expansion in powers of $\sigma^2$ around Eq.~\eqref{eq:sigma0-for-rhonu}, as carried out in Appendices~\ref{app:analytical-derivation} and \ref{app:robustness-cdw}---see Eqs.~\eqref{eq:CDW-x-pos-neg} and \eqref{eq:CDW-x-pos-neg-hi-order}. Therein, we showed that these additional terms still allow for the emergence of periodic solutions with period $a$, but restrict their shape to the cosine and sine CDW modes in Eq.~\eqref{eq:rhonu-cos-sin}. Note that the shape of the CDW modes so obtained does not depend on the potential $U(x)$, due to our assumption \eqref{eq:exp-factor-small} that allowed us to neglect the terms proportional to $e^{-\beta {\Delta U(x \pm a,x)}}$ in Eqs.~\eqref{eq:sigma0-for-rhonu-a} and \eqref{eq:sigma0-for-rhonu-c}, and analogous terms in the higher-order terms in our expansion in powers of $\sigma^2$.

\bibliography{PCPyT24.bib}

\end{document}